\documentclass[superscriptaddress,11pt,nofootinbib,standalone]{revtex4-1}

\usepackage{bbm,bm,graphicx,mathtools,color,hyperref,slashed}
\usepackage[T1]{fontenc}
\usepackage{enumerate}
\usepackage[all]{xy}
\usepackage{amsmath,amssymb,amsthm}
\usepackage{ascmac}
\usepackage{braket}
\usepackage{extarrows}
\usepackage{bm}
\usepackage{subcaption}
\usepackage{here}
\usepackage{cases}
\usepackage{comment}
\usepackage{tikz,tikz-3dplot}
\usetikzlibrary{intersections, calc}

\newcommand{\disT}{{}^{\ast}T^{D}}
\newcommand{\disB}{{}^{\ast}B^{D}}
\newcommand{\disC}{{}^{\ast}T^{D}\times {}^{\ast}B^{d}}

\newcommand{\hmu}{\hat{\mu}}
\newcommand{\U}[1]{U_{n,#1}}
\newcommand{\betti}[2]{\beta_{#1}(#2)}
\newcommand{\calD}{\mathcal{D}}
\newcommand{\pbc}{{\rm PBC}}
\newcommand{\dbc}{{\rm DBC}}
\newcommand{\calU}{\mathcal{U}}
\newcommand{\calV}{\mathcal{V}}
\newcommand{\calW}{\mathcal{W}}
\newcommand{\diag}[1]{\mathrm{Diag} \begin{pmatrix} #1 \end{pmatrix}}

\newtheorem{conj}{Conjecture.}

\begin{document}

\title{
New conjecture on exact Dirac zero-modes of lattice fermions
}

\author{Jun Yumoto}
\email{d8521007(at)s.akita-u.ac.jp}
\address{Department of Mathematical Science, Akita University, 1-1 Tegata-Gakuen-machi, Akita 010-8502, Japan}

\author{Tatsuhiro Misumi}
\email{misumi(at)phys.kindai.ac.jp}
\address{Department of Physics, Kindai University, 3-4-1 Kowakae, Higashi-osaka, Osaka 577-8502, Japan}
\address{Research and Education Center for Natural Sciences, Keio University, 4-1-1 Hiyoshi, Yokohama, Kanagawa 223-8521, Japan}

\begin{abstract}
We propose a new conjecture on the relation between the exact Dirac zero-modes of free and massless lattice fermions and the topology of manifold on which the fermion action is defined. 
Our conjecture claims that the maximal number of exact Dirac zero-modes of fermions on finite-volume and finite-spacing lattices defined by discretizing torus, hyperball, their direct-product space, and hypersphere is equal to the summation of the Betti numbers of their manifolds if several specific conditions on lattcie formulations are satisfied. 
We start with reconsidering exact Dirac zero-modes of naive fermions on the lattices whose topologies are torus, hyperball and their direct-product space ($T^{D} \times B^{d}$).
We find that the maximal number of exact zero-modes of free Dirac fermions is in exact agreement with the sum of Betti numbers $\displaystyle \sum^{D}_{r=0} \beta_{r}$ for these manifolds. Indeed, the $4D$ lattice fermion on torus has up to $16$ zero-modes while the sum of Betti numbers of $T^4$ is $16$. This coincidence holds also for the $D$-dimensional hyperball and their direct-product space $T^{D} \times B^{d}$.
We study several examples of lattice fermions defined on a certain discretized hypersphere ($S^{D}$), and find that it has up to $2$ exact zero-modes, which is the same number as the sum of Betti numbers of $S^{D}$.
From these facts, we conjecture the equivalence of the maximal number of exact Dirac zero-modes and the summation of Betti numbers under the specific conditions. We discuss a program for proof of the conjecture in terms of Hodge theory and spectral graph theory.
\end{abstract}

\maketitle

\newpage

\tableofcontents

\newpage


\section{Introduction}
\label{sec:Intro}

Quantum field theories are non-perturbatively investigated by use of the numerical Monte-Carlo simulation based on lattice field theory \cite{Wilson:1974sk, Creutz:1980zw}.  
One of the most delicate problems in the application is how to introduce fermionic degrees of freedom \cite{Karsten:1980wd, Nielsen:1980rz, Nielsen:1981xu, Nielsen:1981hk}, where we encounter serious problems such as the doubling problem, the difficulty of a single Weyl fermion, and the sign problem of the quark determinant. 
The lattice fermion formulations including Wilson fermions \cite{Wilson:1975id}, Domain-wall or overlap fermions \cite{Kaplan:1992bt, Shamir:1993zy, Furman:1994ky, Neuberger:1998wv, Ginsparg:1981bj}, and staggered fermions \cite{Kogut:1974ag, Susskind:1976jm, Kawamoto:1981hw,Sharatchandra:1981si,Golterman:1984cy,Golterman:1985dz,Kilcup:1986dg} have been proposed to bypass the problems and have been broadly used in the lattice simulation. Apart from them, relatively new approaches have been proposed, including the generalized Wilson fermions \cite{Bietenholz:1999km,Creutz:2010bm,Durr:2010ch,Durr:2012dw, Misumi:2012eh,Cho:2013yha,Cho:2015ffa,Durr:2017wfi},
the staggered-Wilson fermions \cite{Golterman:1984cy, Adams:2009eb,Adams:2010gx,Hoelbling:2010jw, deForcrand:2011ak,Creutz:2011cd,Misumi:2011su,Follana:2011kh,deForcrand:2012bm,Misumi:2012sp,Misumi:2012eh,Durr:2013gp,Hoelbling:2016qfv,Zielinski:2017pko}, the minimally doubled fermion \cite{Karsten:1981gd,Wilczek:1987kw,Creutz:2007af,Borici:2007kz,Bedaque:2008xs,Bedaque:2008jm, Capitani:2009yn,Kimura:2009qe,Kimura:2009di,Creutz:2010cz,Capitani:2010nn,Tiburzi:2010bm,Kamata:2011jn,Misumi:2012uu,Misumi:2012ky,Capitani:2013zta,Capitani:2013iha,Misumi:2013maa,Weber:2013tfa,Weber:2017eds,Durr:2020yqa} 
and the central-branch Wilson fermion \cite{Kimura:2011ik,Creutz:2011cd,Misumi:2012eh,Chowdhury:2013ux}.

The non-torus lattice fermion formulation is also an interesting avenue, where we consider lattices with nonzero-genus topologies. In the previous work of ours, we introduced spectral graph theory to clarify the number of zero-eigenvalues of lattice Dirac operators \cite{YM2022}. By use of this tool, we made progress in understanding the number of exact Dirac zero-modes on lattices with arbitrary topologies and no discrete translational symmetry\footnote{In the other context, the Dirac-Kahler fermions on generic manifolds have been investigated, where the relation between the fermion zero-modes and the Euler characteristics of the manifold are discussed \cite{Catterall:2018lkj,Butt:2021brl}.}.

In this work we propose a new conjecture on the relation between the topology of a manifold and the number of Dirac zero-modes of free lattice fermions defined on a lattice-discretized version of the manifold.  
The claim of our conjecture is that the maximal number of exact Dirac zero-modes of free and massless fermions on a finite-volume and finite-spacing lattice defined by discretizing a $D$-dimensional torus, hyperball, their direct-product space, and hypersphere is equal to the summation of the Betti numbers of the manifolds $\displaystyle \sum_{r=0}^{D} \beta_{r}({\mathcal M})$ as long as the formulation has several specific properties, including locality, $\gamma_{D+1}$-hermiticity\footnote{We define $\gamma_{D+1} \equiv \gamma_1\gamma_2\cdots\gamma_D$ in $D$-dimensional lattice.} and other characteristics such as the discretization manner of sphere.
The $r$-th Betti number is defined as the rank of the $r$-th homology group $H_{r}({\mathcal M})$.
We show that our conjecture holds for lattice fermions on torus ($T^{D}$), hyperball ($B^D$) and their product space $T^{D} \times B^{d}$.
We also investigate lattice fermions defined on specifically discretized hyperspheres ($S^{2}$) and find an empirical fact that they have up to $2$ zero-modes at least for the discretization, while the sum of Betti numbers for $S^{D}$ is $2$ irrespective of dimensions. This explicit investigation indicates that the conjecture holds even for spherical and non-hypercubic lattices for special cases. In the end of this paper, we discuss the way how to prove this coincidence in terms of Hodge theory \cite{Rabin:1981qj, Hod41, Eck45, Dod76, DP76} and spectral graph theory \cite{west2001introduction,bondy1976graph,mieghem_2010,Watts1998,YM2022}. We there introduce a new viewpoint that the lattice Wilson term corresponds to a graph Laplacian operator giving Betti numbers.

The reason we use the terminology, ``the maximal number of exact Dirac zero-modes",  is two-folded:
The number of zero-modes depends on the number of lattice sites when we consider a finite-volume and finite-spacing lattice. Thus, we focus just on the case that the number of zero-modes are ``maximal" in this sense given a lattice-discretized manifold.
It also depends on lattice fermion formulations. Thus, we focus on the case of the ``maximal" number of zero-modes given a certain lattice, which is usually realized by adopting the naive fermion formulation.

We also note that the number of exact Dirac zero modes of free naive fermion on discretized torus lattices is directly related to the number of species (doublers). The argument of this work is thus rephrased in terms of the species doubling for some cases.

This paper is constructed as follows:
In Sec.~\ref{sec:max_doublers}, we present empirical evidences for our conjecture.
In Sec.~\ref{sec:new_conj}, we propose the main conjecture.
In Sec.~\ref{sec:basis_conj}, we discuss a program for the proof of the conjecture.
Sec.~\ref{sec:SD} is devoted to the summary and discussion.


\section{Maximal number of Dirac zero-modes on discretized manifolds}
\label{sec:max_doublers}

In this section, we present empirical evidences that the maximal number of Dirac zero-modes on the discretized manifold is equal to the summation of the Betti numbers of the continuum manifold.
We firstly focus on three types of manifolds, $D$-dimensional torus $T^{D}$, $D$-dimensional hyperball $B^{D}$, and ($D + d$)-dimensional cylinder $T^{D} \times B^{d}$. We also present explicit calculations of zero-modes on the discretized sphere $S^{D}$.


\subsection{$D$-dimensional torus $T^{D}$}
\label{subsec:torus}

\begin{figure}[htpb]
\centering
	\includegraphics[clip,
		height=4cm]{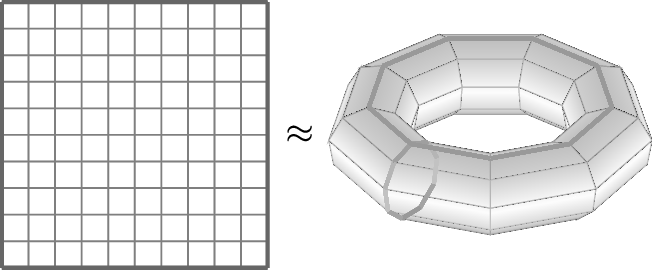}
\caption{Two-dimensional square lattice with opposite boundaries identified corresponds to the discretized two-dimensional torus.
}
\label{fig1}
\end{figure}

The discretized $D$-dimensional torus is realized as the $D$-dimensional square (hypercubic) lattice with the periodic boundary condition (\pbc) imposed in each dimension. Hereafter, we denote the discretized $D$-dimensional torus as $\disT$ in order to distinguish it from the continuum manifold $T^{D}$. The discretized torus in two dimensions is schematically depicted in Fig.~\ref{fig1}.

The free lattice naive-fermion action on $\disT$, or on $D$-dimensional square lattice with \pbc, is
\begin{equation}
\label{eq:torus_act}
    S_{T^{D}} = \sum_{n}\sum_{\mu=1}^{D} 
    \bar{\psi}_{n} \gamma_{\mu} D_{\mu}^{(\pbc)} \psi_{n}
    \equiv \bar{\bm{\psi}} \calD^{(\pbc)} \bm{\psi}\,,
\end{equation}
where $D_{\mu} \equiv (T_{+\mu} - T_{-\mu})/2$ with $T_{\pm\mu}\psi_{n} = U_{n,\pm\hmu}\psi_{n\pm\hmu}$ and $\hmu$ is a unit vector. In a free theory, we just set $\U{\pm\hmu} = \bm{1}$. Unless otherwise specified, we consider a free theory hereafter. The sum $\sum_{n}$ stands for the summation over lattice sites $n = (n_{1}, n_{2},\cdots, n_{D}) \in \mathbb{N}^{D}$, whose intervals are $1 \leq n_{\mu} \leq N$. The vector $\bm{\psi}$ is defined as $\bm{\psi} \equiv \sum_{n} \psi_{n} \ket{n}$ with $\ket{n} \equiv {\displaystyle \bigotimes_{\mu=1}^{D}} \ket{n_{\mu}} = \ket{n_{1}} \otimes \ket{n_{2}} \otimes \cdots \otimes \ket{n_{D}}$. Then, the diagonalized Dirac matrix $\calD$ is obtained as
\begin{equation}
\label{eq:torus_diag}
    \calU^{\dagger} \calD^{(\pbc)} \calU = 
    \sum_{k} \sum_{\mu}^{D} i \sin \left[ \frac{2\pi}{N} \left( k_{\mu} - 1 \right) \right] \gamma_{\mu}
    \ket{k}\bra{k}\,,
\end{equation}
where $k=\left( k_{1}, k_{2} ,\cdots, k_{D} \right) \in [1,N]^{D}$ with $k_{\mu} \in \mathbb{Z}$. The symbol $\calU$ is the unitary matrix defined as $\calU \equiv { \displaystyle \sum_{n,k}} \left\{ \prod_{\mu} \exp \left[ \frac{2\pi i}{N}\left( n_{\mu}-1 \right) \left( k_{\mu}-1 \right) \right] \right\}\ket{n}\bra{k}$ with $\ket{k} \equiv { \displaystyle \bigotimes_{\mu=1}^{D}} \ket{k_{\mu}}$. 
The condition that the Dirac matrix has zero-eigenvalues (zero-modes) is simply given by
\begin{equation}
\label{eq:torus}
    \sum_{\mu}^{D} i\gamma_{\mu} \sin \left[ \frac{2\pi}{N} \left( k_{\mu} - 1 \right) \right] = 0
    \quad \Longrightarrow \quad
    \sin \left[ \frac{2\pi}{N} \left( k_{\mu} - 1 \right) \right] = 0\,,
\end{equation}
as a consequence of the linear independence of $\gamma$-matrices. It shows that we have up to $2^{D}$ Dirac zero-modes (fermion species) on $\disT$ as long as we take an even $N$. For an odd $N$, the number of Dirac zero-modes can decrease, but never increases.
Even if we adopt another fermion formulation such as Wilson and staggered fermions on the discretized torus, but the number of Dirac zero-modes for these cases are certainly smaller than $2^D$. Thus, we argue that the maximal number of Dirac zero-modes is $2^D$ on the discretized torus.
Note that we only consider a finite-volume and finite-spacing lattice here.
We never take a continuum or a thermodynamic limit in this paper.

Meanwhile, the $r$-th Betti number $\betti{r}{T^{D}}$ of the $D$-dimensional torus $T^{D}$ is obtained as
\begin{equation}
\label{eq:Betti_torus}
    \betti{r}{T^{D}} = {}_{D}C_{r}\,,
\end{equation}
where $r$ is an integer from 1 to $D$.
For instance, Betti numbers of four-dimensional torus $T^{4}$ are $\betti{0}{T^{4}} = 1,\ \betti{1}{T^{4}} = 4,\ \betti{2}{T^{4}} = 6,\ \betti{3}{T^{4}} = 4,\ \betti{0}{T^{4}} = 1$.
Then, the summation of the Betti number in Eq.~(\ref{eq:Betti_torus}) over $r$ is 
\begin{equation}
\label{eq:sum_betti}
    \sum_{r=0}^{D} \betti{r}{T^{D}} = \sum_{r=0}^{D} {}_{D}C_{r} = 2^{D}.
\end{equation}
From these results, we argue that the summation of the Betti numbers for the $D$-dimensional torus $T^{D}$ is equal to the maximal number of exact Dirac zero-modes of free fermions on the discretized $D$-dimensional torus $\disT$.


\subsection{$D$-dimensional hyperball $B^{D}$}

The discreized $D$-dimensional hyperball $\disB$ is realized as the $D$-dimensional square lattice with Dirichlet boundary condition (\dbc) imposed in each dimension.
The difference from $\disT$ is that there is a pair of unconnected boundaries in each direction. The free lattice naive-fermion action on $\disB$ is
\begin{equation}
    S_{B^{D}} = \sum_{n}\sum_{\mu=1}^{D} 
    \bar{\psi}_{n} \gamma_{\mu} D_{\mu}^{(\dbc)} \psi_{n}
    \equiv \bar{\bm{\psi}} \calD^{(\dbc)} \bm{\psi}.
\end{equation}
Note that this action is different from Eq.~(\ref{eq:torus_act}) in that there are no hoppings between the boundaries. The diagonalized Dirac matrix by the unitary matrix $\calV \equiv \sum_{n,k} \left( \prod_{\mu} i^{k_{\mu}}\sin \left[ \frac{n_{\mu}k_{\mu}\pi}{N+1} \right] \right) \ket{n}\bra{k}$ is 
\begin{equation}
\label{eq:ball_diag}
    \calV^{\dagger} \calD^{(\dbc)} \calV = 
    \sum_{k} \sum_{\mu}^{D} i \cos \left[ \frac{k_{\mu}\pi}{N+1} \right] \gamma_{\mu}
    \ket{k}\bra{k}\,,
\end{equation}
where $k=\left( k_{1}, k_{2} ,\cdots, k_{D} \right) \in [1,N]^{D}$ with $k_{\mu} \in \mathbb{Z}$. The Dirac matrix has zero-eigenvalues (zero-modes) when Eq.~(\ref{eq:torus_diag}) satisfies
\begin{equation}
\label{eq:ball}
    \sum_{\mu}^{D} i\gamma_{\mu} \cos \left[ \frac{k_{\mu}\pi}{N+1} \right] = 0
    \quad \Longrightarrow \quad
    \cos \left[ \frac{k_{\mu}\pi}{N+1} \right] = 0.
\end{equation}
It means that there is one Dirac zero-mode, or equivalently equivalently a single zero-mode, if we take an odd number as $N$. Eq.~(\ref{eq:ball}) has no solution since $k_{\mu} = \frac{N+1}{2}$ is not an integer if $N$ is even. 
As with the case of torus, other fermion formulations never increase the number zero-modes. Hence, we argue that the maximal number of Dirac zero-modes on $\disB$ is one. 
We also note that we just consider the bulk fermionic degrees of freedom, not an edge mode.

On the other hand, the $r$-th Betti numbers for the hyperball are given as
\begin{equation}
\label{eq:Betti_ball}
    \betti{r}{B^{D}} = \delta_{r0}\,,
\end{equation}
where $\delta$ stands for the Kronecker delta. 
The summation of the Betti numbers in Eq.~(\ref{eq:Betti_ball}) is 
\begin{equation}
    \sum_{r=0}^{D} \betti{r}{B^{D}} 
    = \sum_{r=0}^{D} \delta_{r0} 
    = 1.
\end{equation}
From these results, we claim that the maximal number of Dirac zero-modes of free fermions on $D$-dimensional lattice-discretized hyperball $\disB$ is equal to the summation of the Betti numbers over $0\leq r \leq D$ for the $D$-dimensional hyperball $B^{D}$.

If we take an infinite-volume limit, the number of naive Dirac zero-modes on $D$-dimensional discretized hyperball $\disB$ approaches $2^{D}$, which is the same as that on $D$-dimensional discretized torus.
This number is usually called ``the number of species".
It is notable that {\it our conjecture on exact Dirac zero modes is applicable only to the finite-volume lattice.}

\subsection{($D+d$)-dimensional product manifold $T^{D} \times B^{d}$}

\begin{figure}[htpb]
\centering
	\includegraphics[clip,
		height=4cm]{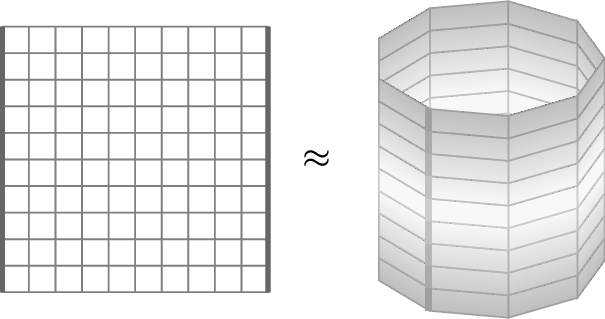}
\caption{($1+1$)-dimensional square lattice with the left and right boundaries connected is identified as the discretized hollow cylinder.
}
\label{fig2}
\end{figure}

The ($D+d$)-dimensional product manifold $T^{D} \times B^{d}$ consists of two manifolds, the $D$-dimensional torus $T^{D}$ and the $d$-dimensional hyperball $B^{d}$. 
It means that the discretized ($D+d$)-dimensional product manifold $\disC$ consists of the discretized $D$-dimensional torus $\disT$ and the discretized $d$-dimensional hyperball $\disB$. 
The ($D+d$)-dimensional square lattice is identified as $\disC$ by imposing PBC on the $D$-dimensions and imposing DBC on the remaining $d$-dimensions. 
The manifold in ($1+1$)-dimensions is depicted in Fig.~\ref{fig2}.

The free lattice naive-fermion action on this lattice is 
\begin{equation}
    S_{T^{D} \times B^{d}} = \sum_{n} \bar{\psi}_{n}
    \left[
    \sum_{\mu=1}^{D} \gamma_{\mu}D_{\mu}^{(\pbc)}
    +
    \sum_{\nu=D+1}^{d} \gamma_{\nu}D_{\nu}^{(\dbc)}
    \right] \psi_{n}
    \equiv \bar{\bm{\psi}} 
        \calD^{(\mathrm{P}+\mathrm{D})}
    \bm{\psi}\,.
    \label{cylac}
\end{equation}
We note that, if we add the Wilson term to this action in ($4+1$) dimensions, it results in the action of Domain-wall fermion. For this case, the zero-modes of the Dirac matrix emerge as edge modes.
The Dirac matrix $\calD^{(\mathrm{P}+\mathrm{D})}$ in Eq.~(\ref{cylac}) can be diagonalized by the unitary matrix $\calW \equiv \sum_{k} \left\{ \prod_{\mu} \exp \left[ \frac{2\pi i}{N}\left( n_{\mu}-1 \right) \left( k_{\mu}-1 \right) \right] \right\}\left\{ \prod_{\nu} i^{k_{\nu}}\sin \left[ \frac{n_{\nu}k_{\nu}\pi}{N+1} \right] \right\} \ket{n}\bra{k}$ with $\ket{n} \equiv \bigotimes_{\lambda=1}^{D+d} \ket{n_{\lambda}}$ and $\ket{k} \equiv \bigotimes_{\lambda=1}^{D+d} \ket{k_{\lambda}}$. The diagonalized Dirac matrix is
\begin{equation}
\label{eq:cyli_diag}
    \calW^{\dagger} \calD^{(\mathrm{P}+\mathrm{D})} \calW
    = i \left\{
    \sum_{\mu=1}^{D} \gamma_{\mu} \sin \left[ \frac{2\pi}{N} \left( k_{\mu} - 1 \right) \right]
    +
    \sum_{\nu=D+1}^{D+d} \gamma_{\nu} \cos \left[ \frac{k_{\nu}\pi}{N+1} \right]
    \right\}\ket{k}\bra{k}\,.
\end{equation}
The condition for Eq.~(\ref{eq:cyli_diag}) to have zero eigenvalues (zero-modes) is
\begin{equation}
\label{eq:cilinder}
\begin{gathered}
    \sum_{\mu=1}^{D} \gamma_{\mu} \sin \left[ \frac{2\pi}{N} \left( k_{\mu} - 1 \right) \right]
    +
    \sum_{\nu=D+1}^{D+d} \gamma_{\nu} \cos \left[ \frac{k_{\nu}\pi}{N+1} \right]
    = 0\\
    \quad \Longrightarrow \quad
    \sin \left[ \frac{2\pi}{N} \left( k_{\mu} - 1 \right) \right] 
    = 0, \quad
    \cos \left[ \frac{k_{\nu}\pi}{N+1} \right]
    = 0\,.
\end{gathered}
\end{equation}
It shows that the maximal number of zero-modes on $\disC$ is $2^{D} = 2^{D} \times 1$, which is the product of the maximum numbers of Dirac zero-modes on $\disT$ and $\disB$. The other fermion formulations, including the Domain-wall setup, never increase the number of zero-modes, but can decrease it. 
 
The Betti numbers of the product manifold are 
\begin{equation}
     \beta_{r}(T^{D} \times B^{d}) = 
     \begin{cases}
        {}_{D}C_{r} &(1 \leq r \leq D)\\
        0           &(r > D)
     \end{cases}\,.
\end{equation}
The summation of the $r$-th Betti numbers ($0\leq r \leq D+d$) for the ($D+d$)-dimensional product manifold $T^{D} \times B^{d}$ is
\begin{equation}
    \sum_{r=0}^{D+d} \beta_{r}(T^{D} \times B^{d})
    = \sum_{r=0}^{D} {}_{D}C_{r} = 2^{D}.
\end{equation}
These results indicate that the maximum number of Dirac zero-modes of free fermions on $\disC$ is equal to the summation of the Betti numbers of $T^{D} \times B^{d}$.

We also note that the maximal number of naive Dirac zero-modes approaches $2^{D+d}$ in an infinite-volume limit. 
Our statement of the conjecture holds only for finite-volume cases.


\subsection{$D$-dimensional sphere $S^{D}$}

In the continuum field theory, the fermion action on spheres gives massive fermionic degrees of freedom since the curvature works as effective mass. It is, however, not the case on the discretized sphere.

In this subsection, we study the number of exact Dirac zero-modes on the discretized sphere, where we perform discretization and put a fermion on the lattice in a special manner. We empirically show that the maximal number of Dirac zero-modes on the discretized sphere is equal to the sum of the Betti number of $D$-dimensional sphere $S^{D}$ at least for the discretization.

We begin with the two-dimensional cases.
We firstly consider the following discretized spherical coordinate system for $2$-sphere, labeled by two integers $(M,N)$:
\begin{equation}
    x_{3} = r\cos \theta_{2},
    \qquad
    x_{2} = r\sin \theta_{2} \cos \theta_{1},
    \qquad
    x_{1} = r\sin \theta_{2} \sin \theta_{1},
\end{equation}
where $r$ is a radial distance and two angles are discretized as $\theta_{2} \equiv \left( N-n_{2} \right)\pi/ \left( N-1 \right),\ \theta_{1} \equiv 2n_{1}\pi/M$ with $n_{1},n_{2} \in \mathbb{N}$ and $n_{1} \in [1,N],\ n_{2} \in [1,M]$.
For simplicity, we fix a radial distance as $r=1$.
We label lattice sites as $(n_{1},n_{2})$.
Note that there are two special points $(n_{1},1)$,  $(n_{1},N)$ who ignore the hopping in $n_{1}$-direction.
We call the two points the south pole, relabeled as $(0,1)$, and the north pole, relabeled as $(0,N)$, respectively.
By this definition, we consider the naive-fermion-like action on the discretized 2-sphere labeled by $(M,N)$ as
\begin{equation}
    S_{(M,N)} = 
    \sum_{n} \sum_{\mu=1}^{2} \bar{\psi}_{n} \sigma_{\mu}D_{\mu} \psi_{n}
    \equiv \bm{\bar{\psi}} \mathcal{D}^{(M,N)} \bm{\psi}\,,
\end{equation}
where $\bm{\psi}$ is a vector defined as $\bm{\psi} = \sum_{n} \psi_{n}\bm{e}_{n}$ with the standard basis $\bm{e}_{n}$ and $\mathcal{D}^{(M,N)}$ is the Dirac matrix.
The sum $\sum_{n}$ stands for the summation over lattice site $n = (n_{1},n_{2})$.
Here, we specify the order of components $\psi_{n}$ in the vector as $(0,1) \to (1,2) \to (2,2) \to \cdots \to (M,2) \to (1,3) \to \cdots \to (M,N-1) \to (0,N)$. Namely, it is expressed as
\begin{equation}
    \bm{\psi} = 
    \begin{pmatrix}
        \psi_{(0,1)}\\
        \psi_{(1,2)}\\
        \psi_{(2,2)}\\
        \vdots\\
        \psi_{(M,2)}\\
        \psi_{(1,3)}\\
        \vdots\\
        \psi_{(M,N-1)}\\
        \psi_{(0,N)}
    \end{pmatrix}\,.
\end{equation}
The Dirac matrix $\mathcal{D}^{(M,N)}$ is expressed as
\begin{equation}
\label{eq:sphere_action}
	\mathcal{D}^{(M,N)} 
	= \frac{1}{2}
	\begin{pmatrix}
		I_{N-2} \otimes P_{M}	\\
						&O_{2}
	\end{pmatrix} \otimes \sigma_{1} 
	+ \frac{1}{2}
	\begin{pmatrix}
	D_{N-2} \otimes I_{M}		&V_{M(N-2),2}\\
	-V_{M(N-2),2}^{\dagger}	&O_{2}
	\end{pmatrix} \otimes \sigma_{2}\,,
\end{equation}
where $\sigma_{i}$ is the Pauli matrix and $I_{k}$ and $O_{k}$ are the $k$ dimensional identity matrix and null matrix, respectively.
$P_{M}$ is the $M \times M$ difference matrix with the periodic boundary condition and $D_{N-2}$ is the $(N-2) \times (N-2)$ difference matrix with the Dirichlet boundary condition as
\begin{equation}
	P_{M} = 
	\begin{pmatrix}
	0	&1		&0	&		&0	&0		&-1	\\
	-1	&0		&1	&\cdots	&0	&0		&0	\\
	0	&-1		&0	&		&0	&0		&0	\\
		&\vdots	&	&\ddots	&	&\vdots		\\
	0	&0		&0	&		&0	&1		&0	\\
	0	&0		&0	&\cdots	&-1	&0		&1	\\
	1	&0		&0	&		&0	&-1		&0	
	\end{pmatrix},	\quad
	D_{N-2} =
	\begin{pmatrix}
	0	&1		&0	\\
	-1	&0		&1	\\
	0	&-1		&0	\\
		&		&	&\ddots	\\
		&		&	&		&0	&1		&0	\\
		&		&	&		&-1	&0		&1	\\
		&		&	&		&0	&-1		&0
	\end{pmatrix}.
\end{equation}
$V_{M(N-2),2}$ is given by
\begin{equation}
	V_{M(N-2),2} \equiv 
	\begin{pmatrix}
		1	&0\\
		0	&0\\
		 \multicolumn{2}{c}{\vdots}\\
		 0	&0\\
		 0	&-1
	\end{pmatrix}_{N-2,2} \otimes
	\begin{pmatrix}
		1\\
		1\\
		 \vdots\\
		 1\\
		 1
	\end{pmatrix}_{M,1}\,,
\end{equation}
where the subscripts $j,k$ of the matrix $(\ )_{j,k}$ stand for the row and column sizes.

\begin{figure}[htpb]
\begin{minipage}[t]{0.45\linewidth}
\centering
	\includegraphics[clip,
		height=4cm
		]{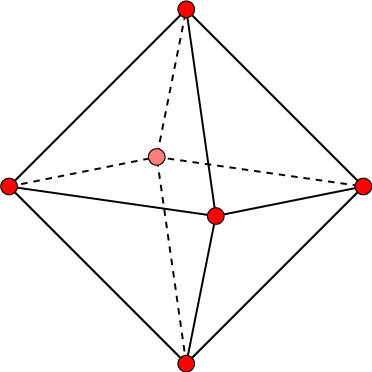}
\subcaption{Discretized 2-sphere $(4,3)$. 
}
\label{graph:43}
\end{minipage}
\begin{minipage}[t]{0.45\linewidth}
\centering
	\includegraphics[clip,
		height=4cm
		]{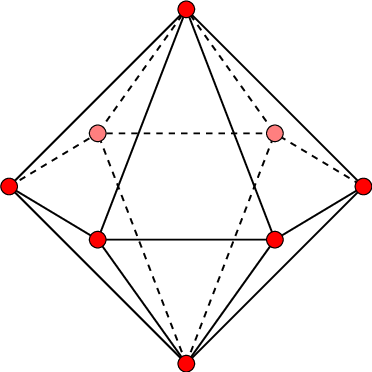}
\subcaption{Discretized 2-sphere $(6,3)$. 
}
\label{graph:63}
\end{minipage}
\caption{
The discretized 2-spheres $(4,3)$ and $(6,3)$ are depicted.
Red points stand for lattice sites and lines between the red points stand for links.
}
\label{graph:tri2sph}
\end{figure}

In the example of the discretized 2-sphere $(M,3)$, the action is
\begin{equation}
    S_{(M,3)} = \bm{\bar{\psi}} \mathcal{D}^{(M,3)} \bm{\psi}\,.
\end{equation}
The discretized 2-spheres $(4,3)$ and $(6,3)$ are depicted in Fig.~\ref{graph:tri2sph}.
As seen from the figures, these cases correspond to a sort of the triangulation of 2-sphere.

The Dirac matirx $\mathcal{D}^{(M,3)}$ is
\begin{equation}
    \mathcal{D}^{(M,3)} 
    = \frac{1}{2} 
    \begin{pmatrix}
        P_{M}\\ &   O_{2}
    \end{pmatrix} \otimes \sigma_{1}
    + \frac{1}{2}
    \begin{pmatrix}
        I_{M}   &V_{M,2}\\
        -V_{M,2}^{\dagger}  &O_{2}
    \end{pmatrix} \otimes \sigma_{2}\,,
\end{equation}
where $V_{M,2}$ is 
\begin{equation}
    V_{M,2} = 
    \begin{pmatrix}
        1   &-1
    \end{pmatrix}_{1,2}
    \otimes
    \begin{pmatrix}
		1\\
		1\\
		 \vdots\\
		 1\\
		 1
	\end{pmatrix}_{M,1}\,.
\end{equation}
This matrix is diagonalized by the unitary matrix
\begin{equation}
    \mathcal{U}^{(M,3)} \equiv U^{(M,3)} \otimes I_{2}\,,
\end{equation}
where $I_{2}$ is a $2 \times 2$ identity matrix and $U^{(M,3)}$ is 
\begin{equation}
\begin{split}
    &U^{(M,3)} = \frac{1}{\sqrt{M}}\\
    &\times\begin{pmatrix}
        \sqrt{\frac{M}{2}}	&0		&0		&0		&\cdots		&0			&0			&-\sqrt{M}|\chi|^{2}		&-\sqrt{M}|\chi|^{2}		\\
		0	&1		&1		&1		&\cdots		&1		&1		&\chi		&\bar{\chi}	\\
		0	&\xi		&\xi^{2}		&\xi^{3}		&\cdots		&\xi^{M-2}		&\xi^{M-1}		&\chi		&\bar{\chi}	\\
		0	&\xi^{2}	&\xi^{4}		&\xi^{6}		&\cdots		&\xi^{2(M-2)}	&\xi^{2(M-1)}	&\chi		&\bar{\chi}	\\
		\vdots &\vdots	&\vdots		&\vdots		&\ddots		&\vdots		&\vdots		&\vdots	&\vdots	\\
		0	&\xi^{M-3}	&\xi^{2(M-3)}	&\xi^{3(M-3)}	&\cdots		&\xi^{(M-3)(M-2)}	&\xi^{(M-3)(M-1)}		&\chi		&\bar{\chi}	\\
		0	&\xi^{M-2}	&\xi^{2(M-2)}	&\xi^{3(M-2)}	&\cdots		&\xi^{(M-2)^{2}}		&\xi^{(M-2)(M-1)}		&\chi		&\bar{\chi}	\\
		0	&\xi^{M-1}	&\xi^{2(M-1)}	&\xi^{3(M-1)}	&\cdots		&\xi^{(M-1)(M-2)}		&\xi^{(M-1)^{2}}		&\chi		&\bar{\chi}	\\
		\sqrt{\frac{M}{2}}	&0		&0		&0		&\cdots		&0			&0			&\sqrt{M}|\chi|^{2}		&\sqrt{M}|\chi|^{2}
    \end{pmatrix}
\end{split}\,,
\end{equation}
with $\xi \equiv e^{\frac{2\pi}{M}\left( n_{1}-1 \right)}$ and $\chi \equiv i/\sqrt{2}$.
Then, the spectra of the Dirac matrix $\mathcal{D}^{(M,3)}$ is obtained as
\begin{equation}
\label{eq:diag_2sphere}
\begin{split}
    &\left( U^{(M,3)} \right)^{\dagger} \calD^{(M,3)} U^{(M,3)} \\
		&= \diag{
			0,
			&\sin \left[ \frac{2\pi}{M} \right] \sigma_{1},
			&\sin \left[ \frac{4\pi}{M} \right] \sigma_{1},
			&\cdots,
			&\sin\left[ \frac{2\pi \left( M-1 \right) }{M} \right] \sigma_{1},
			&-i \sqrt{\frac{M}{2}} \sigma_{2},
			&i \sqrt{\frac{M}{2}} \sigma_{2}
		}
\end{split}\,.
\end{equation}
From Eq.~(\ref{eq:diag_2sphere}), one finds that the number of Dirac zero-modes of the Dirac matrix is algebraically two for even $M$ since there is a certain $j$ satisfying $j =\frac{M}{2} + 1 \in \mathbb{N}$ and $\sin \left[ \frac{2\pi \left( j-1 \right) }{M} \right] = 0$ other than $j=1$.
On the other hand, the number of Dirac zero-modes is one for odd $M$ since only $j=1$ satisfies $\sin \left[ \frac{2\pi \left( j-1 \right) }{M} \right] = 0$ for this case.

These results indicate that there are up to two zero-modes on the discretized 2-sphere $(M,3)$.
We show that the maximal number of zero-modes are two also in other cases including $(M,N)=(4,4), (5,4), (5,5), (4,5), (6,9)$ by numerical calculations as shown in Appendix.~\ref{app:sphere1}. The results are summarized in Table.~\ref{table:2sphere}.

The maximal number of exact Dirac zero-modes, ``two", for this discretization with the naive-fermion-like setup is equal to the sum of the Betti numbers for the 2-sphere
\begin{equation}
    \sum_{r=0}^{2} \beta_{r}(S^{2}) 
    = \sum_{r=0}^{2} \left( \delta_{r0} + \delta_{r2} \right)
    = 2\,,
\end{equation}
where the $r$-th Betti number for the 2-sphere is $\beta_{r}(S^{2}) = \delta_{r0} + \delta_{r2}$.

In higher dimensions, we discuss the naive-fermion-like action on the cellular-decomposed sphere in a parallel manner.
We show that there are up to two exact zero-modes on the discretized 4-sphere labeled by four integers $(N_{1},N_{2},N_{3},N_{4})$ by numerical calculation in Appendix.~\ref{app:sphere2}.
On the other hand, the $r$-th Betti numbers ($0\leq r \leq D$) for $D$-dimensional sphere $S^{D}$ is given as
\begin{equation}
    \beta_{r}(S^{D}) = \delta_{r0} + \delta_{rD}\,.
\end{equation}
Thus, the sum of the Betti numbers for the $D$-dimensional sphere is given as
\begin{equation}
    \sum_{r=0}^{D} \beta_{r}(S^{D}) 
    = \sum_{r=0}^{D} \left( \delta_{r0} + \delta_{rD} \right)
    = 2\,.
\end{equation}
From these results, we argue that the maximal number of Dirac zero-modes of the naive-fermion-like setup on the $D$-dimensional discretized sphere ${}^{\ast}S^{D}$ is equal to the sum of the Betti numbers for the $D$-dimensional sphere $S^{D}$ at least for this discretization.

It is notable that the lattice fermion action on the spherical lattice has been studied in the literature in the different context \cite{Kamata:2016xmu, Kamata:2016rqi, Brower2017}. Our result is consistent with the observations obtained in the literature.

In the end of this section, we make a comment on possible zero-mode difference giving the quantum anomaly in gravitational background.
In curved space, the difference of the numbers of left-handed and right-handed zero-modes are related to the anomaly resulting from the gravitational background.
Our argument of the existence of two exact zero modes is not inconsistent to this difference of the numbers of zero-modes. 
For example, the case with two right-handed zero-modes and zero left-handed zero-modes are consistent with both arguments. 
This kind of fixing of zero modes may be due to the specific choice of the sphere discretization or due to the lattice artifact, but there is no contradiction so far.


\section{New conjecture}
\label{sec:new_conj}

In the previous section, we have discussed that the maximal number of Dirac zero modes of lattice fermions in terms of the Betti numbers of the manifold. In Table.~\ref{table:betti}, we summarize the relation of the sum of the Betti numbers and the maximal number of Dirac zero-modes.
\begin{table}[t]
\caption{Betti numbers and Maximal numbers of Dirac zero-modes}
\label{table:betti}
\begin{tabular}{c|c|c}
    \hline
    \quad manifold $M$ \quad
    &\quad sum of $\beta_{r}(M)$ \quad
    &\quad maximal $\#$ of Dirac zero-modes \quad \\ \hline\hline
    \quad $1$-d torus \quad
    &\quad $1+1$ \quad
    &\quad $2$ \quad \\ \hline
    \quad $2$-d torus \quad
    &\quad $1+2+1$ \quad
    &\quad $4$ \quad \\ \hline
    \quad $3$-d torus \quad
    &\quad $1+3+3+1$ \quad
    &\quad $8$ \quad \\ \hline
    \quad $4$-d torus \quad
    &\quad $1+4+6+4+1$ \quad
    &\quad $16$ \quad \\ \hline
    \quad Torus $T^{D}$ \quad
    &\quad $\left( 1+1 \right)^{D}$ \quad
    &\quad $2^{D}$ \quad \\ \hline
    \quad Hyperball $B^{D}$ \quad
    &\quad $1+0+0+\cdots$ \quad
    &\quad $1$ \quad \\ \hline
    \quad Sphere $S^{D}$ \quad
    &\quad $1+0+0+\cdots+1$ \quad
    &\quad $2$ \quad \\ \hline
    \quad $T^{D}\times B^{d}$ \quad
    &\quad $2^{D}\times 1$ \quad
    &\quad $2^{D}$ \quad \\ \hline
\end{tabular}
\centering
\end{table}

From these facts, we propose a new conjecture on the number of exact Dirac zero-modes on the discretized torus, hyperball, their direct-product space, and hypersphere. Hereafter, we denote these manifolds as $\mathcal M$.
The conjecture is as follows:
\begin{conj}
We firstly impose the following six conditions on the free fermion action of $\mathcal M$:
\begin{enumerate}[i.]
    \item
    Difference operator; we adopt a certain difference in Dirac operator so that anti-hermiticity of the Dirac matrix in the action holds.
    For the torus, it is just a central difference.
    It is not a central difference at the boundary of the hyperball since we take the Dirichlet boundary condition for it.
    For the case of sphere lattices, we adopt a difference similar to the central difference as discussed in Eq.~\ref{eq:sphere_action}. 
    \item
    $\gamma_{D+1}$ hermiticity or axis-symmetric Dirac spectrum; we only consider lattice fermions with real-axis-symmetric Dirac eigenvalue spectrum. This condition is satisfied by $\gamma_{D+1}$ hermiticity in even dimensions.
    \item
    $2^{D/2}$ or $2^{(D+1)/2}$ spinors; this condition assures the linear independence of the lattice action for each direction.
    When $D$ is even(odd), we consider $2^{D/2}$($2^{(D+1)/2}$) spinors.
    \item
    Locality; this condition leads to finite-hopping actions although it may be unnecessary for our conjecture because non-locality usually decreases the number of zero-modes.
    \item
    Finite volume lattice; our conjecture claims that the fermion action on the finite-volume lattice picks up the topology of the continuum manifold.
    \item
    Fermion formulation on sphere; we restrict our argument on spheres only to the specific fermion formulation we adopt in this paper. We here do not introduce spin connection or vierbein since the fermion is defined on the spherical coordinate.
\end{enumerate}
Our conjecture claims that, as long as these conditions hold, the maximal number of exact Dirac zero-modes of free fermions on the lattice-discretized $D$-dimensional manifold is equal to the summation of Betti numbers $\beta_{r}({\mathcal M})$ over $0\leq r\leq D$ for the continuum manifold ${\mathcal M}$.
It is expressed as
\begin{equation}
 {\rm max}[ {\mathcal N}({}^{\ast}{\mathcal M}) ]\,=\, \sum_{r=0}^{D} \beta_{r}({\mathcal M})\,,
\end{equation}
where ${\mathcal N}({}^{\ast}{\mathcal M})$ is the number of exact Dirac zero-modes on the lattice-discretized manifold ${}^{\ast}{\mathcal M}$. 
\end{conj}


\section{Program for proof}
\label{sec:basis_conj}

In the previous section, we propose the conjecture on the equivalence between the maximal number of exact Dirac zero-modes of free fermions on the discretized manifold and the summation of Betti numbers of the continuum manifold, by studying the $D$-dimensional torus-lattice ${}^{\ast}T^{D}$, the hyperball-lattice ${}^{\ast}B^{D}$, the $D \times d$ dimensional cylinder-lattice ${}^{\ast}T^{D} \times {}^{\ast}B^{d}$ and the $D$-dimensional sphere lattice ${}^{\ast}S^{D}$.
We now propose a program for proof of the conjecture in terms of spectral graph theory and Hodge theory.
In Hodge theory \cite{Hod41, Eck45, Dod76, DP76}, {\it the number of zero-eigenvalues of a $r$-th Laplacian ($\Delta_{r} = \partial_{r+1}\partial_{r+1}^{*} + \partial_{r}^{*} \partial_{r}$) defined on a complex chain coincides with the $r$-th Betti number}, where $\partial_{r}$ is a $r$-th boundary operator.
Our program for the proof is inspired by this fact.

In our program, we re-interpret lattice fermions as spectral graphs \cite{YM2022}. We define a Laplacian operator $\mathcal{L}$ of a graph $G(V,E)$ having the set of vertices $V$ and the set of edge $E$ as
\begin{equation}
    \mathcal{L}_{ij} \equiv \begin{cases}
        d_{i}   &\mbox{if $i = j$}\\
        -1  &\mbox{if $i \neq j$ and $(i,j)$ are linked}\\
        0   &\mbox{if $i \neq j$ and $(i,j)$ are not linked}
    \end{cases}\,,
    \label{eq:Lij}
\end{equation}
where $d_{i}$ is the number of edges sharing the site $i$.
Vertices in graph theory correspond to lattice sites in lattice field theory while edges correspond to links. The detailed correspondence between lattice field theory and spectral graph theory is summarized in our previous work \cite{YM2022}.


\subsection{One-dimensional lattices}

We begin with one-dimensional lattices. 
We consider two types of one-dimensional lattices: the one with PBC ($1$-dim discretized torus ${}^{\ast}T^{1}$) and the other with DBC ($1$-dim discretized hyperball ${}^{\ast}B^{1}$).

For the one-dimensional lattice with $N$ vertices with PBC, the diagonal elements of Laplacian matrix $d_i$ are $d_i = 2$ for $i=1,2,...,N$, thus the Laplacian in Eq.~(\ref{eq:Lij}) is 
\begin{equation}
\label{eq:Lap}
    \mathcal{L}_{ij}
    = 2\sum_{k=1}^{N} \delta_{ik}\delta{kj}
        - \left( \delta_{iN}\delta_{1j} + \delta_{i1}\delta_{Nj} 
        + \sum_{k=1}^{N-1}\delta_{ik}\delta_{k+1\,j}
        + \sum_{k=2}^{N}\delta_{ik}\delta_{k-1\,j}\right)\,.
\end{equation}
For simplicity, we take an even $N$ for this case.
It is notable that the number of zero-eigenvalues of this Laplacian is one, which can be checked by direct calculation.
From the generic argument on graph Laplacian matrices in topological graph theory, it is known that the number of zero-eigenvalues of $\mathcal{L}$ of the graph $G(V,E)$ represents the number of connected components, or equivalently the $0$-th Betti number $\beta_{0}(G)$ of $G(V,E)$.
Since an certain manifold in one dimension and its lattice-discretized graph share the same Betti numbers ($\beta_{0}, \beta_{1}$), the number of zero-eigenvalues of $\mathcal{L}$ in Eq.~(\ref{eq:Lap}) gives the $0$-th Betti number of one-dimensional torus. Indeed, the $0$-th Betti number of one-dimensional torus is one.

The Lagrangian constructed from the Laplacian matrix and the field vector $\bm{\psi} = \sum_{i=1}^{N} \psi_{i}\bm{e}_{i}$ results in the Wilson term $S_{\mathrm{Wt}}$ with the mass parameter $m$ set to zero and the Wilson-fermion parameter $r$ set to one in a free theory, where it is expressed as
\begin{equation}
\label{eq:Wf}
\begin{split}
    \bm{\bar{\psi}} \mathcal{L} \bm{\psi}
    &= \sum_{n=1}^{N} \bar{\psi}_{n} \left[ \psi_{n+1} + \psi_{n-1} -2 \psi_{n} \right]\\
    &= 2\sum_{n=1}^{N} \bar{\psi}_{n} 
    \left. \left[
        m\psi_{n} - \frac{r}{2} \left( \psi_{n+1} + \psi_{n-1} -2 \psi_{n} \right)
    \right] \right|_{m=0,\, r=1}
    \equiv \left. S_{\mathrm{Wt}} \right|_{m=0,\, r=1}
\end{split}\,.
\end{equation}
This fact means that the number of zero-eigenvalues of the matrix corresponding to the Wilson term on ${}^{\ast}T^{1}$ is equal to the $0$-th Betti number $\beta_{0}=1$ for the continuum torus $T^{1}$.
We will later discuss that it also holds for ${}^{\ast}B^{1}$ and $B^{1}$.

We next consider another Laplacian operator defined as
\begin{equation}
    \mathcal{L}'_{ij} \equiv \begin{cases}
        -d_{i}  &\mbox{if $i = j$}\\
        -1      &\mbox{if $i \neq j$ and $(i,j)$ are linked}\\
        0       &\mbox{if $i \neq j$ and $(i,j)$ are not linked}
    \end{cases}\,,
    \label{eq:Lapp}
\end{equation}
with $d_{i}=2$ on ${}^{\ast}T^{1}$.
This Laplacian corresponds to the case with the mass parameter $m = -2$ in Eq.~(\ref{eq:Wf}).
The number of zero-eigenvalues of this Laplacian $\mathcal{L}'$ on ${}^{\ast}T^{1}$ is one, which is equivalent to the first Betti number of the continuum torus $T^{1}$.
From this fact, we obtain one observation that {\it the number of zero-eigenvalues of $\mathcal{L}'$ gives the first Betti number in one dimension}, which we have to prove in the future.

To convince readers, we study the one-dimensional lattice with $N$ vertices with DBC (${}^{\ast}B^{1}$), where the diagonal elements of the Laplacian matrices $d_i$ are $d_1 = 1$, $d_N = 1$, and $d_i = 2$ for $i=2,...,N-1$, where we take an odd $N$ for this case.
For the Laplacian $\mathcal{L}$ in Eq.~(\ref{eq:Lap}), the number of zero-eigenvalues of $\mathcal{L}$, which again corresponds to the Wilson term on ${}^{\ast}B^{1}$, gives the number of connected components, thus it is gives the $0$-th Betti number $\beta_{0}=1$ of $B^{1}$.
For the Laplacian $\mathcal{L}'$ in Eq.~(\ref{eq:Lapp}),
the number of its zero-eigenvalues on ${}^{\ast}B^{1}$ is zero, which is equal to the first Betti number of $B^{1}$.
It is notable that our observation on the equivalence of the number of zero eigenvalues of $\mathcal{L}'$ and the first Betti number also holds for ${}^{\ast}B^{1}$ and $B^{1}$.

So far, we study the Laplacian corresponding to the Wilson term on the one-dimensional lattice (graph) and find the following facts:
\begin{itemize}

\item
The number of zero-eigenvalues of $\mathcal{L}$ corresponding to the Wilson term with $m=0, r=1$ is equal to $\beta_{0}$ of the corresponding manifold.
It is a mathematically rigorous fact.

\item
The number of zero-eigenvalues of $\mathcal{L}'$ corresponding to the Wilson term with $m=-2, r=1$ seems to be equal to $\beta_{1}$ of the corresponding manifold.

\end{itemize}

Two zero-eigenvectors of the free naive Dirac matrix (Dirac operator) on ${}^{\ast}T^{1}$, which are found from the condition in Eq.~(\ref{eq:torus}), are identical to the one of $\mathcal{L}$ and the one of $\mathcal{L}'$, respectively, while one zero-eigenvector of the naive Dirac matrix on ${}^{\ast}B^{1}$, which is found from the condition in Eq.~(\ref{eq:ball}), is identified with that of $\mathcal{L}'$.
These results originate in the fact that the naive Dirac matrix and the Laplacian operators ($\mathcal{L}$, $\mathcal{L}'$) are simultaneously diagonalizable.

It strongly indicates that the summation of Betti numbers on an one-dimensional manifold ($T^{1}$ or $B^{1}$) is in exact agreement with the number of zero-eigenvalues of the naive Dirac matrix on the lattice-discretized manifold (${}^{\ast}T^{1}$ or ${}^{\ast}B^{1}$), which gives the maximal number of species.
To complete our proof for one dimension, however, we have to prove that the number of zero-eigenvalues of $\mathcal{L}'$ is $\beta_{1}$ of the corresponding manifold.

\subsection{Higher-dimensional lattices}

For the higher-dimensional torus-lattice ${}^{\ast}T^{D}$, the hyperball-lattice ${}^{\ast}B^{D}$ and the cylinder-lattice ${}^{\ast}T^{D} \times {}^{\ast}B^{d}$, we can utilize Kunneth theorem claiming that the homology groups of two cellular chain complexes $C,C'$ and $C\times C'$ have the following relation:
\begin{align}
H_{r}(C\times C')\,\cong\, \bigoplus_{p+q = r} H_{p}(C)\otimes H_{q}(C') \,.
\end{align}
This theorem means the homology group and its rank (Betti number) of a certain two-dimensional product manifold is obtained from those of the one-dimensional manifolds. 
By repeating this, we can obtain the homology groups and Betti numbers for any higher-dimensional manifolds.

At least for ${}^{\ast}T^{D}$, ${}^{\ast}B^{D}$ and ${}^{\ast}T^{D} \times {}^{\ast}B^{d}$, we consider that we can extend the argument in the previous subsection to higher-dimensional cases. In this extension, the Laplacian operators giving Betti numbers $\beta_{r}$ are expressed as the sum of tensor products of the Laplacians $\mathcal{L}$ in Eq.~(\ref{eq:Lap}) and $\mathcal{L}'$ in Eq.~(\ref{eq:Lapp}).

We now give an example: the Laplacian giving $0$-th Betti number for four-dimensional lattices is expressed as
\begin{equation}
\label{eq:Lap_four}
\begin{split}
    \mathcal{L}^{(D=4)}_{r=0} 
    &= \Bigl(
      \mathcal{L} \otimes \bm{1}_{N} \otimes \bm{1}_{N} \otimes \bm{1}_{N}
    + \bm{1}_{N} \otimes \mathcal{L} \otimes \bm{1}_{N} \otimes \bm{1}_{N}\\
    &\qquad
    + \bm{1}_{N} \otimes \bm{1}_{N} \otimes \mathcal{L} \otimes \bm{1}_{N}
    + \bm{1}_{N} \otimes \bm{1}_{N} \otimes \bm{1}_{N} \otimes \mathcal{L}
    \Bigr) \otimes \bm{1}_{4}
\end{split}\,,
\end{equation}
where the lattice has $N^{4}$ vertices (lattice sites). The symbol $\bm{1}_{N}$ is a $N\times N$ identity matrix.
The Lagrangian constructed from this Laplacian is nothing but the four-dimensional Wilson term with $m=0, r=1$.
The Laplacian giving $4$-th Betti number for four-dimensional lattices is expressed as
\begin{equation}
\label{eq:Lap_four}
\begin{split}
    \mathcal{L}^{(D=4)}_{r=4} 
    &= \Bigl(
      \mathcal{L}' \otimes \bm{1}_{N} \otimes \bm{1}_{N} \otimes \bm{1}_{N}
    + \bm{1}_{N} \otimes \mathcal{L}' \otimes \bm{1}_{N} \otimes \bm{1}_{N}\\
    &\qquad
    + \bm{1}_{N} \otimes \bm{1}_{N} \otimes \mathcal{L}' \otimes \bm{1}_{N}
    + \bm{1}_{N} \otimes \bm{1}_{N} \otimes \bm{1}_{N} \otimes \mathcal{L}'
    \Bigr) \otimes \bm{1}_{4}
\end{split}\,,
\end{equation}
which corresponds to the four-dimensional Wilson term with $m=-8,r=1$.
The Laplacian giving other Betti numbers are also expressed as some tensor product of $\mathcal{L}$ and $\mathcal{L}'$, resulting in a certain Wilson term for which one of the branches of Wilson Dirac spectrum is set to the origin.
As a by-product, this procedure of proof shows that the number of Dirac zero-modes on the well-known five branches of Wilson Dirac spectrum $1,4,6,4,1$ stand for the Betti numbers $\beta_{r}(T^{4})$ for $r=0,1,2,3,4$.

This generalization to higher-dimensions is directly applicable to the manifold obtained as a product space of $T^{D}$ and $B^{D}$. For the manifold such as $S^{D}$, we have to develop a more generic way of generalization to higher-dimensions.

In the end of this section, we comment on another avenue toward proof of the conjecture:
Squaring the free naive Dirac matrix leads to another Laplacian operator.
If we can prove that the number of zero-modes of this Laplacian is the sum of the Betti numbers of the continuum manifold, we can easily give a generic proof for any kind of manifolds including $S^{D}$.


\section{Summary and Discussion}
\label{sec:SD}

Our conjecture claims that the maximal number of exact Dirac zero-modes of free fermions on the finite lattices we formulate in the paper is equal to the summation of the Betti numbers of the $D$-dimensional manifold from which the lattice is constructed.
It is summarized as
\begin{equation}
 {\rm max}[ {\mathcal N}({}^{\ast}{\mathcal M}) ]\,=\, \sum_{r=0}^{D} \beta_{r}({\mathcal M})\,,
\end{equation}
where ${\mathcal N}({}^{\ast}{\mathcal M})$ is the number of Dirac zero-modes on the lattice, which is defined as a lattice-discretized version ${}^{\ast}{\mathcal M}$ of the manifold $\mathcal M$. 
In a sense that this conjecture relates the number of lattice Dirac zero-modes to the topology of spacetime manifold, it is complementary to the Nielsen-Ninomiya's no-go theorem which claims the emergence of pairs of fermion species as a result of the cancellation of chiral charges on torus.

In this paper we restrict our argument to the specific finite lattices; the torus, hyperball and their product lattices are composed of an one-dimensional chain complex with PBC or DBC, while the sphere lattice is constructed in a certain spherical coordinate.
For infinite-volume lattices things drastically change:
For example, the number of zero-modes of naive fermion on the one-dimensional lattice hyperball ${}^{\ast}B^1$ approaches two in an infinite-volume limit, which is the same number as that on ${}^{\ast}T^1$.
It is of great importance that our conjecture relates the topology of {\it a continuum manifold} and the zero-modes on {\it a finite lattice defined by discretizing the manifold}. 

Although our argument is restricted to the specific cases so far,
we consider that the conjecture is valid for lattice fermions defined on more generic lattices. 
If the theorem is established for generic cases, it has impacts on the study of lattice field theory. For example, one can predict the number of exact Dirac zero modes of free fermions on non-standard lattices such as discretized double torus.
Future works will be devoted to generalization and establishment of this conjecture.

Study on the connection between lattice field theory and graph theory leads to very rich understanding on both of them.
In the upcoming work of ours, we will discuss the relation between lattice scalar field theory and topological graph theory, where the massless scalar operator on the lattice is exactly given by the graph Laplacian operator.


\begin{acknowledgements}
T. M. and J. Y. are grateful to S.~Aoki, S.~Matsuura and K.~Ohta for the fruitful discussion on the topics of the paper.
The completion of this work is owed to the discussion in the workshop "Lattice field theory and continuum field theory" at Yukawa Institute for Theoretical Physics, Kyoto University (YITP-W-22-02).  
This work of J. Y. is supported by the Sasakawa Scientific Research Grant from The Japan Science Society.
This work of T. M. is supported by the Japan Society for the Promotion of Science (JSPS) Grant-in-Aid for Scientific Research (KAKENHI) Grant Numbers 19K03817.
\end{acknowledgements}


\appendix
\section{Numerical analysis for $D$-dimensional spheres}

\subsection{Two-dimensional sphere}
\label{app:sphere1}
In this appendix, we show that there are up to two Dirac zero-modes (four eigenvalues) on the discretized 2-sphere labeled by $(M,N)$\footnote{Hereafter, we consider the number of sites on the latitude direction $N$ is $N>3$.}
by numerical calculations:
\begin{itemize}
\item 
If both the number of sites on the longitude direction $M$ and the number of sites on the latitude direction $N$ are even, there is  no Dirac zero-modes of the Dirac matrix.
For instance, we consider the discretized 2-sphere labeled by $(4,4)$, which is depicted in Fig.~\ref{graph:44}.
\begin{figure}[htpb]
\centering
	\includegraphics[clip,
		height=4cm
		]{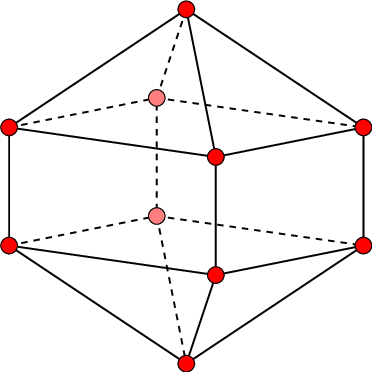}
\caption{Discretized 2-sphere labeled by $(4,4)$.
}
\label{graph:44}
\end{figure}

The Dirac matrix on the $(4,4)$ sphere is
\begin{equation}
    \mathcal{D}^{(4,4)} = \frac{1}{2}
	\begin{pmatrix}
	   	I_{2} \otimes P_{4}	\\
		  &O_{2}
	\end{pmatrix} \otimes \sigma_{1} 
	+ \frac{1}{2}
	\begin{pmatrix}
	   D_{2} \otimes I_{4}		&V_{8,2}\\
	   -V_{8,2}^{\dagger}	&O_{2}
	\end{pmatrix} \otimes \sigma_{2}\,,
\end{equation}
with
\begin{equation}
    V_{8,2} = \begin{pmatrix}
        1   &0\\
        0   &-1
    \end{pmatrix}_{2,2}
    \otimes
    \begin{pmatrix}
        1\\
        1\\
        1\\
        1
    \end{pmatrix}\,.
\end{equation}
The eigenvalues of this Dirac matrix are depicted in Fig.~\ref{plot:2sphere44}.
There is no zero-eigenvalue in $\mathcal{D}^{(4,4)}$, which means that there is no Dirac zero-modes.
\begin{figure}[htpb]
\centering
	\includegraphics[clip,
		width=0.8\linewidth
		]{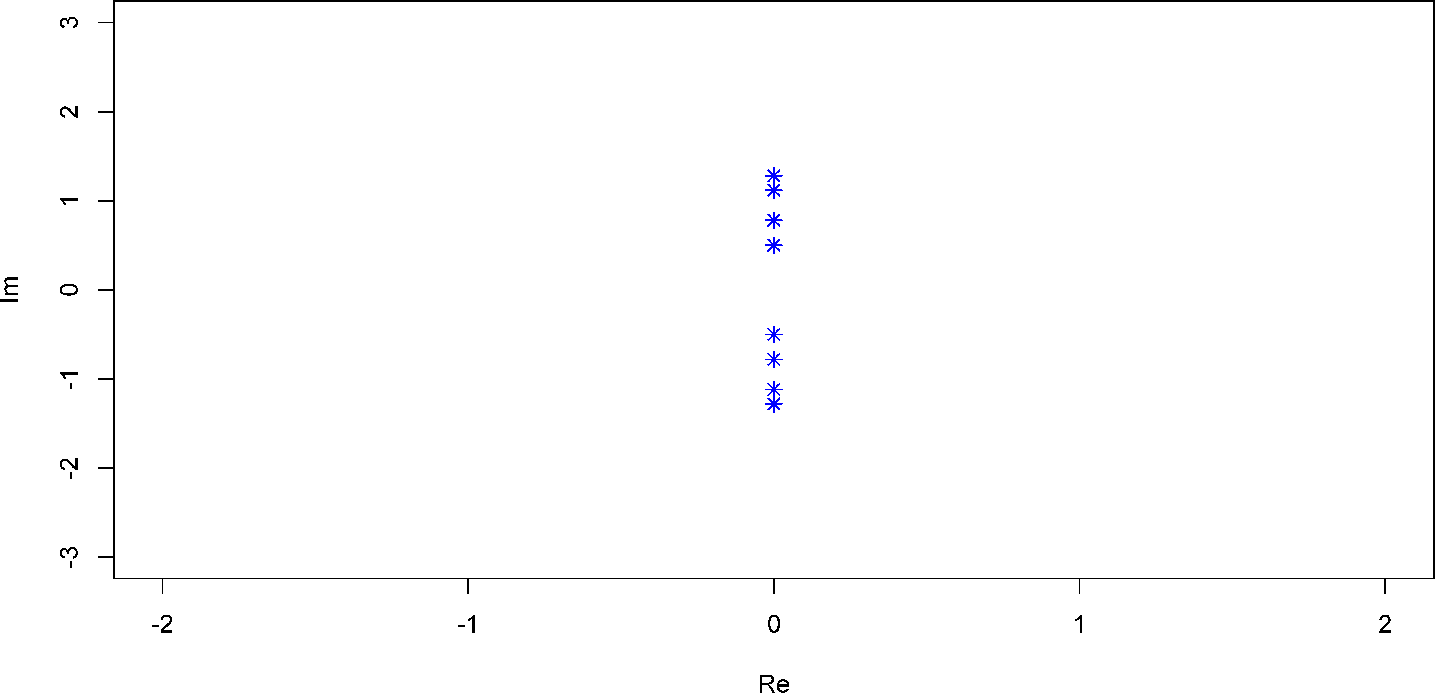}
\caption{Eigenvalue distribution of the Dirac matrix $\mathcal{D}^{(4,4)}$.
There is no Dirac zero-modes, or equivalently no zero-eigenvalue.
}
\label{plot:2sphere44}
\end{figure}

\item
If $M$ is odd and $N$ is even, there is again no Dirac zero-modes (no zero-eigenvalue).
For instance, we take the discretized 2-sphere labeled by $(5,4)$.
We numerically find that the number of zero-modes or the number of Dirac zero-modes is zero as shown in Fig.~\ref{plot:2sphere54}.
\begin{figure}[htpb]
\centering
	\includegraphics[clip,
		width=0.8\linewidth
		]{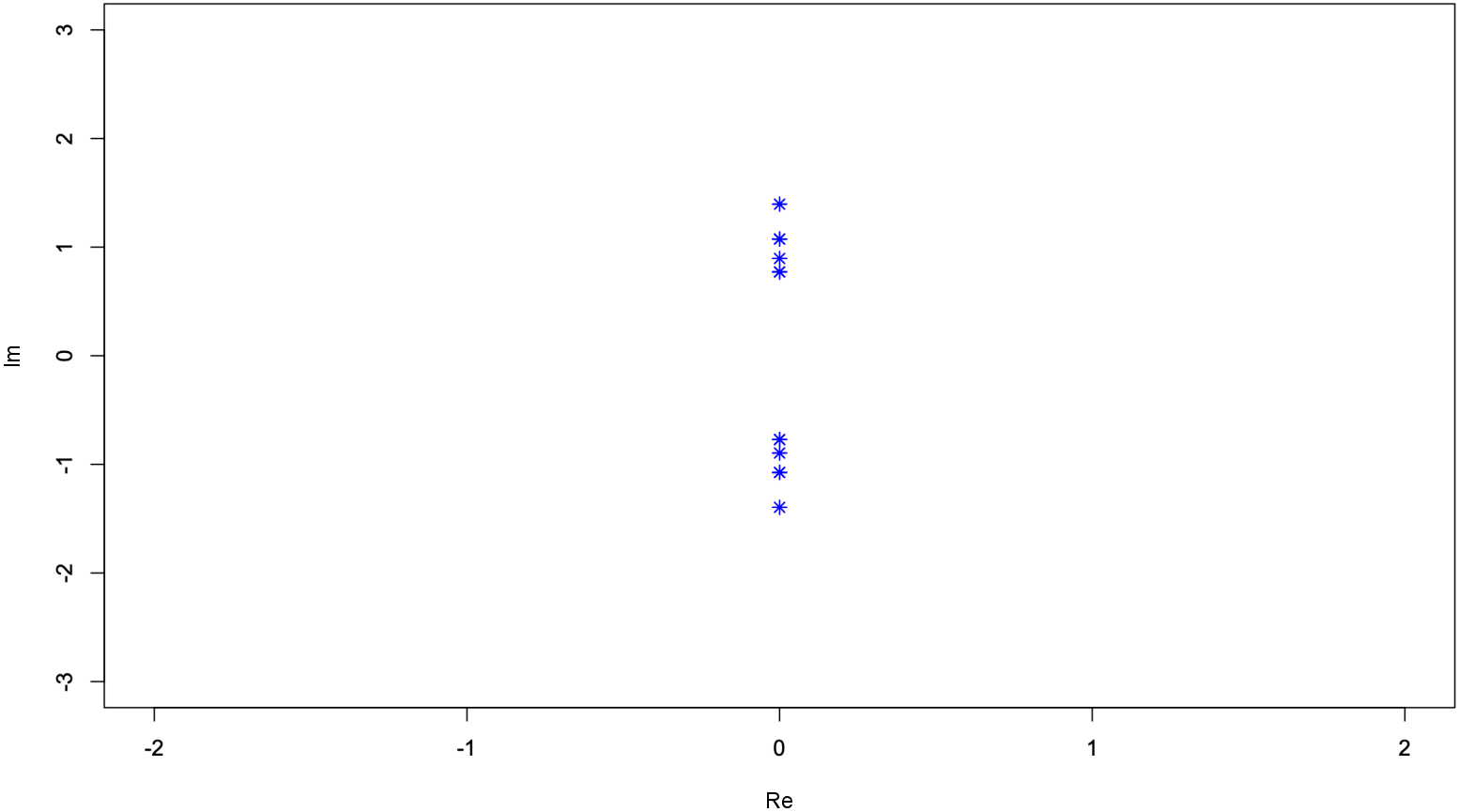}
\caption{
Eigenvalue distribution of the Dirac matrix $\mathcal{D}^{(5,4)}$.
There is no Dirac zero-modes (no zero-eigenvalue) in $\mathcal{D}^{(5,4)}$.
}
\label{plot:2sphere54}
\end{figure}

\item
If both of $M$ and $N$ are odd, there is one Dirac zero-modes (two zero-eigenvalues).
For instance, we consider the discretized 2-sphere labeled by $(5,5)$.
Then, we numerically find that the number of Dirac zero-modes (two zero-eigenvalues) is one as shown in Fig.~\ref{plot:2sphere55} and Fig.~\ref{plot2:2sphere55}.
We note that a pair of eigenvalues corresponds to {\it a single Dirac mode} since the two-dimensional $\gamma$ matrices are $2\times 2$ matrices. Thus, the existence of the pair of zero-eigenvalues shown in Fig.~\ref{plot:2sphere55} and Fig.~\ref{plot2:2sphere55} means that there is a single Dirac zero-mode.
\begin{figure}[htpb]
\centering
	\includegraphics[clip,
		width=0.8\linewidth
		]{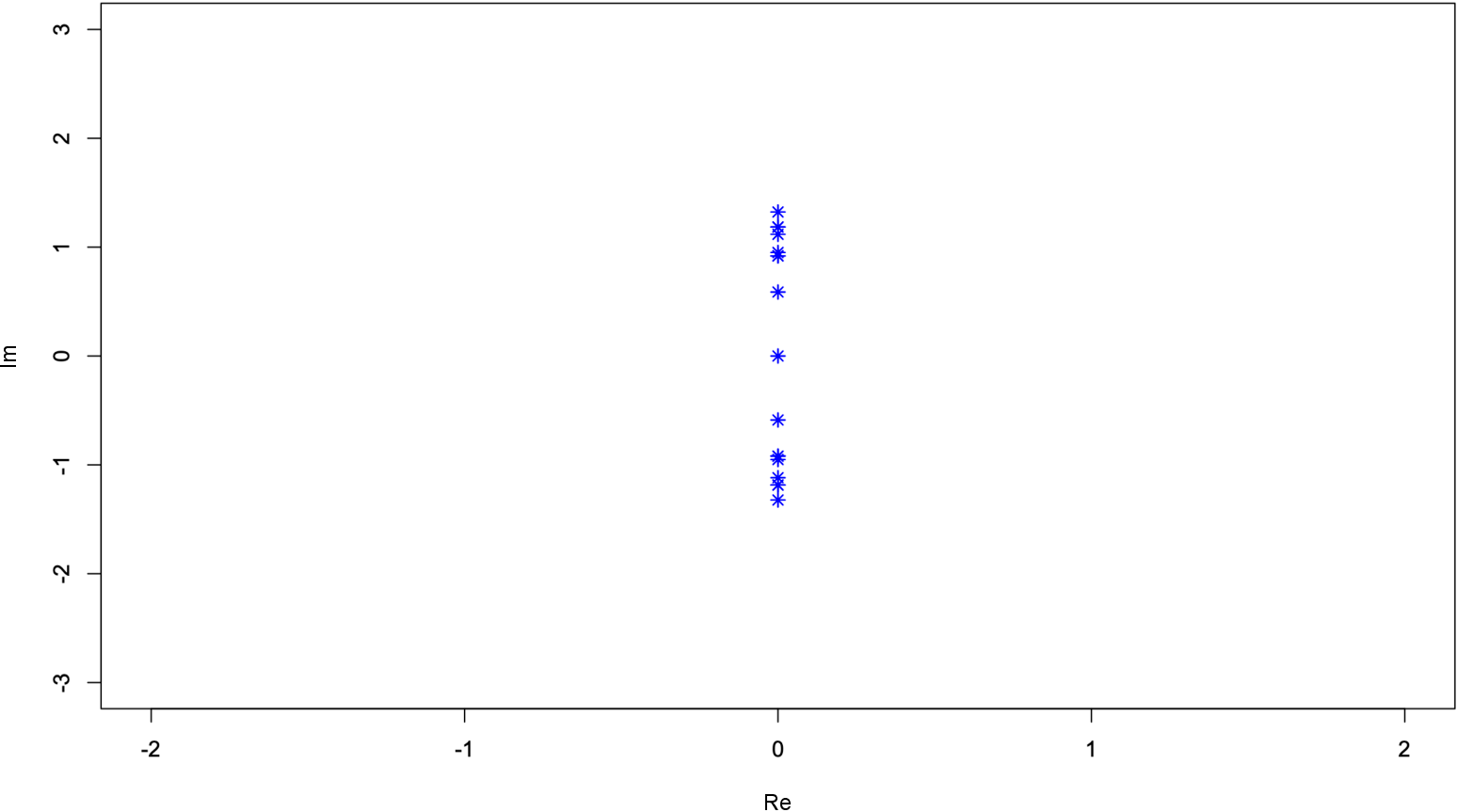}
\caption{
Eigenvalue distribution of the Dirac matrix $\mathcal{D}^{(5,5)}$.
The pair of zero eigenvalues corresponds to a single Dirac zero-mode.
}
\label{plot:2sphere55}
\end{figure}
\begin{figure}[htpb]
\centering
	\includegraphics[clip,
		width=0.8\linewidth
		]{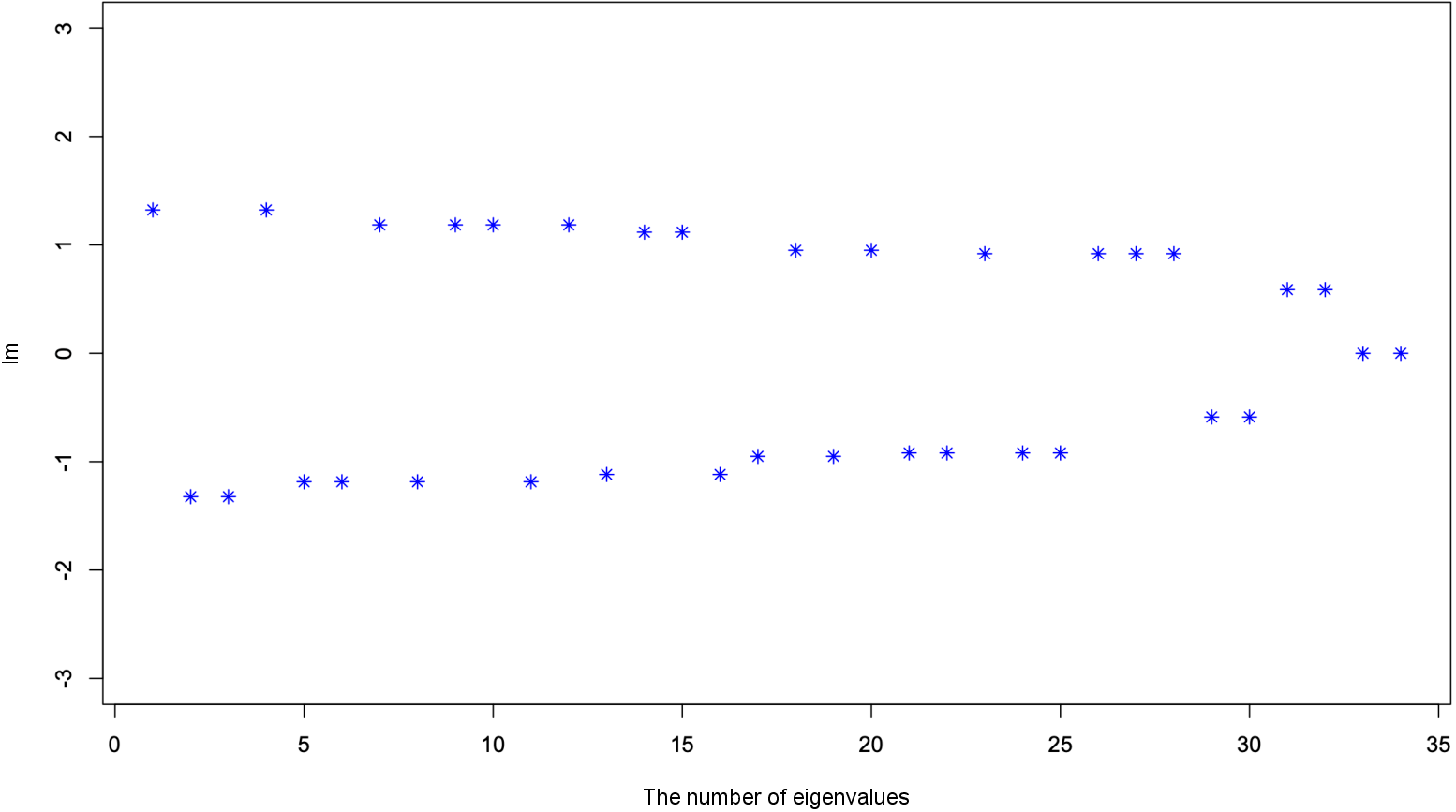}
\caption{
Eigenvalue distribution is depicted, where the vertical axis represents the imaginary part of the Dirac matrix $\mathcal{D}^{(5,5)}$ and the horizontal axis represents the serial number of eigenvalues.
The pair of zero eigenvalues corresponds to a single Dirac zero-mode.
}
\label{plot2:2sphere55}
\end{figure}

\item
If $M$ is even and $N$ is odd, there are two Dirac zero-modes (four eigenvalues) on the discretized 2-sphere.
For instance, we take two cases, $(4,5)$ and $(6,9)$.
The discretized 2-sphere $(4,5)$ is depicted in Fig.~\ref{graph:45}.
\begin{figure}[htpb]
\centering
	\includegraphics[clip,
		height=4cm
		]{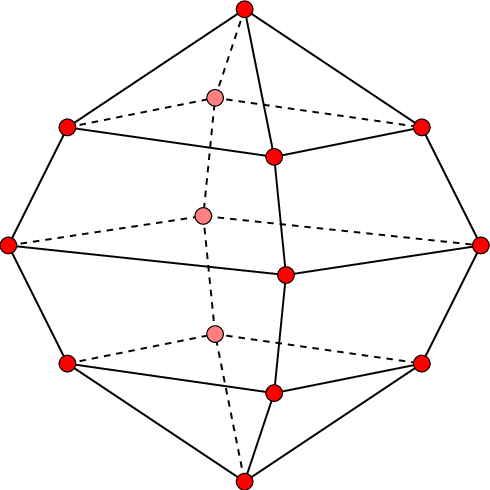}
\caption{Discretized 2-sphere labeled by $(4,5)$.
}
\label{graph:45}
\end{figure}

The Dirac matrix on the $(4,5)$ sphere is
\begin{equation}
    \mathcal{D}^{(4,5)} = \frac{1}{2}
    \begin{pmatrix}
	   	I_{3} \otimes P_{4}	\\
		  &O_{2}
	\end{pmatrix} \otimes \sigma_{1} 
	+ \frac{1}{2}
	\begin{pmatrix}
	   D_{3} \otimes I_{4}		&V_{12,2}\\
	   -V_{12,2}^{\dagger}	&O_{2}
	\end{pmatrix} \otimes \sigma_{2}\,,
\end{equation}
with
\begin{equation}
    V_{12,2} = \begin{pmatrix}
        1  &0\\
        0  &0\\
        0  &-1
    \end{pmatrix}_{3,2}
    \otimes
    \begin{pmatrix}
        1\\
        1\\
        1\\
        1
    \end{pmatrix}\,.
\end{equation}
The eigenvalues of this Dirac matrix are depicted in Fig.~\ref{plot:2sphere45}.
There are two Dirac zero-modes on 2-sphere $(4,5)$ since there are four zero-eigenvalues as seen from Fig.~\ref{plot2:2sphere45}.
\begin{figure}[htpb]
\centering
	\includegraphics[clip,
		width=0.8\linewidth
		]{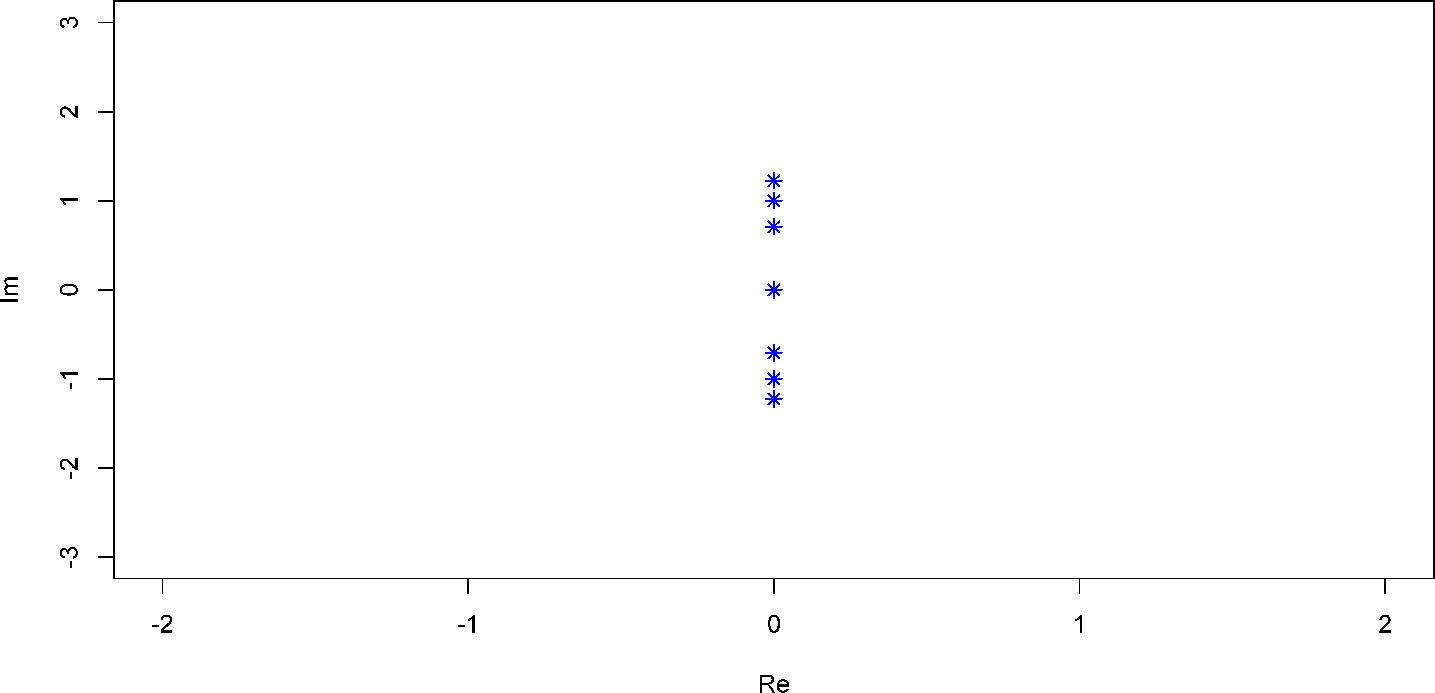}
\caption{Eigenvalue distribution of the Dirac matrix $\mathcal{D}^{(4,5)}$.
There are four zero-eigenvalues corresponding to two Dirac zero-modes in two dimensions.
}
\label{plot:2sphere45}
\end{figure}
\begin{figure}[htpb]
\centering
	\includegraphics[clip,
		width=0.8\linewidth
		]{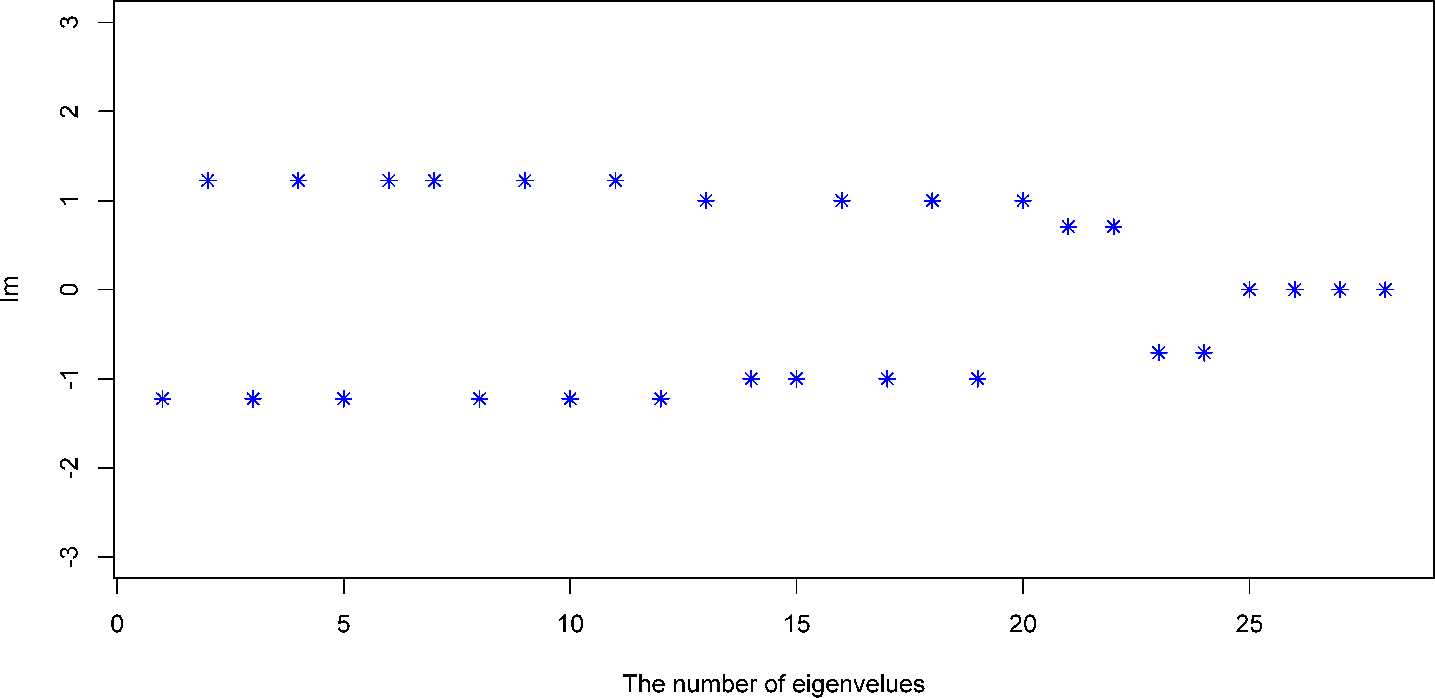}
\caption{Eigenvalue distribution is depicted, where the vertical axis represents the imaginary part of the Dirac matrix $\mathcal{D}^{(4,5)}$ and the horizontal axis represents the serial number of eigenvalues.
There are four zero-eigenvalues corresponding to two Dirac zero-modes in two dimensions.
}
\label{plot2:2sphere45}
\end{figure}

In the case of the $(6,9)$ 2-sphere, the Dirac matrix is
\begin{equation}
    \mathcal{D}^{(6,9)} = \frac{1}{2}
    \begin{pmatrix}
	   	I_{7} \otimes P_{6}	\\
		  &O_{2}
	\end{pmatrix} \otimes \sigma_{1} 
	+ \frac{1}{2}
	\begin{pmatrix}
	   D_{7} \otimes I_{6}		&V_{42,2}\\
	   -V_{42,2}^{\dagger}	&O_{2}
	\end{pmatrix} \otimes \sigma_{2}\,,
\end{equation}
with
\begin{equation}
    V_{42,2} = \begin{pmatrix}
        1  &0\\
        0  &0\\
        0  &0\\
        0  &0\\
        0  &0\\
        0  &0\\
        0  &-1
    \end{pmatrix}_{7,2}
    \otimes
    \begin{pmatrix}
        1\\
        1\\
        1\\
        1\\
        1\\
        1
    \end{pmatrix}\,.
\end{equation}
The eigenvalues of this Dirac matrix are depicted in Fig.~\ref{plot:2sphere69}.
There are again two Dirac zero-modes as there are four zero-eigenvalues in $\mathcal{D}^{(6,9)}$ as seen from from Fig.~\ref{plot2:2sphere69}.
\begin{figure}[htpb]
\centering
	\includegraphics[clip,
		width=0.8\linewidth
		]{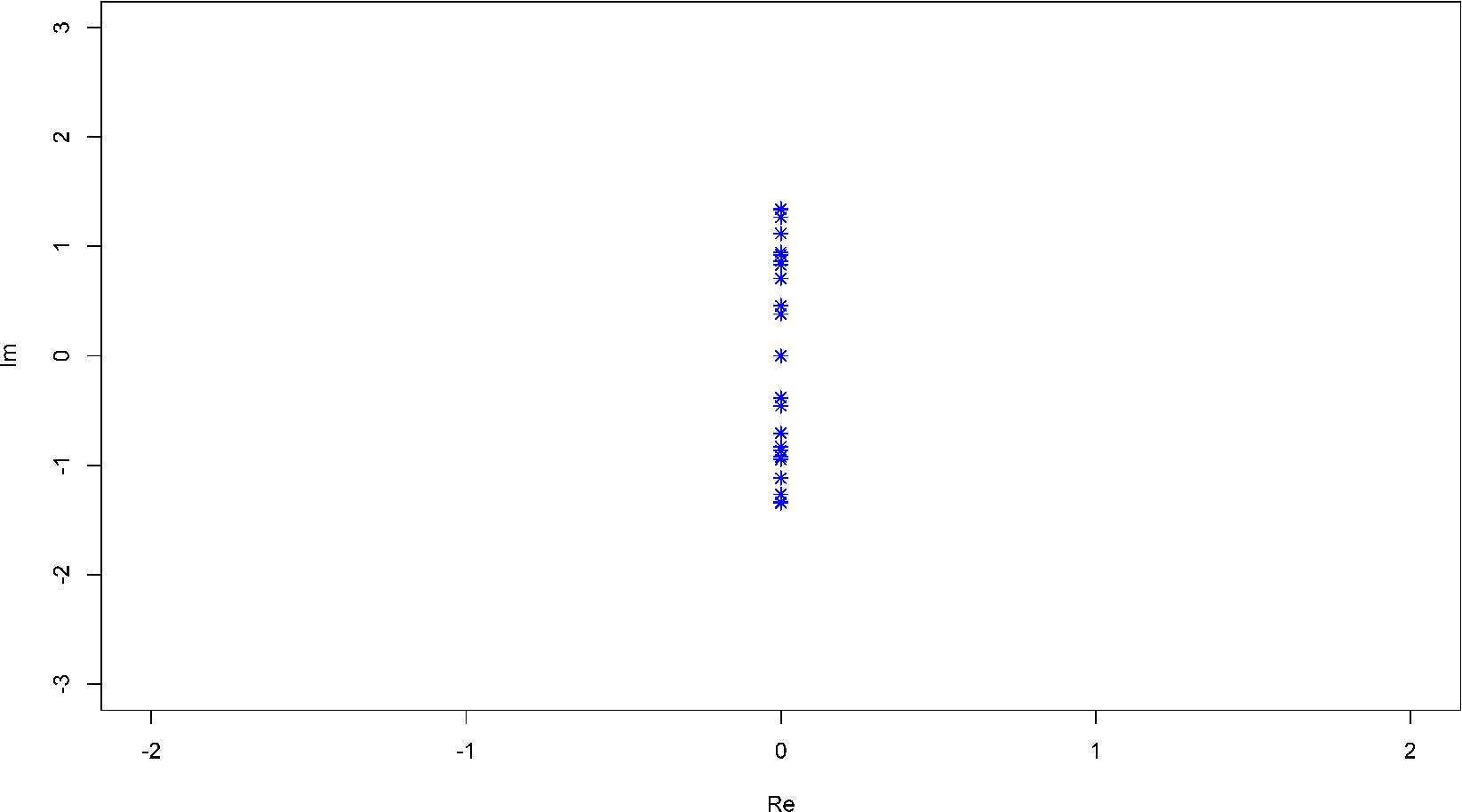}
\caption{Eigenvalue distribution of the Dirac matrix $\mathcal{D}^{(6,9)}$.
There are two Dirac zero-modes (four zero-eigenvalues).
}
\label{plot:2sphere69}
\end{figure}
\begin{figure}[htpb]
\centering
	\includegraphics[clip,
		width=0.8\linewidth
		]{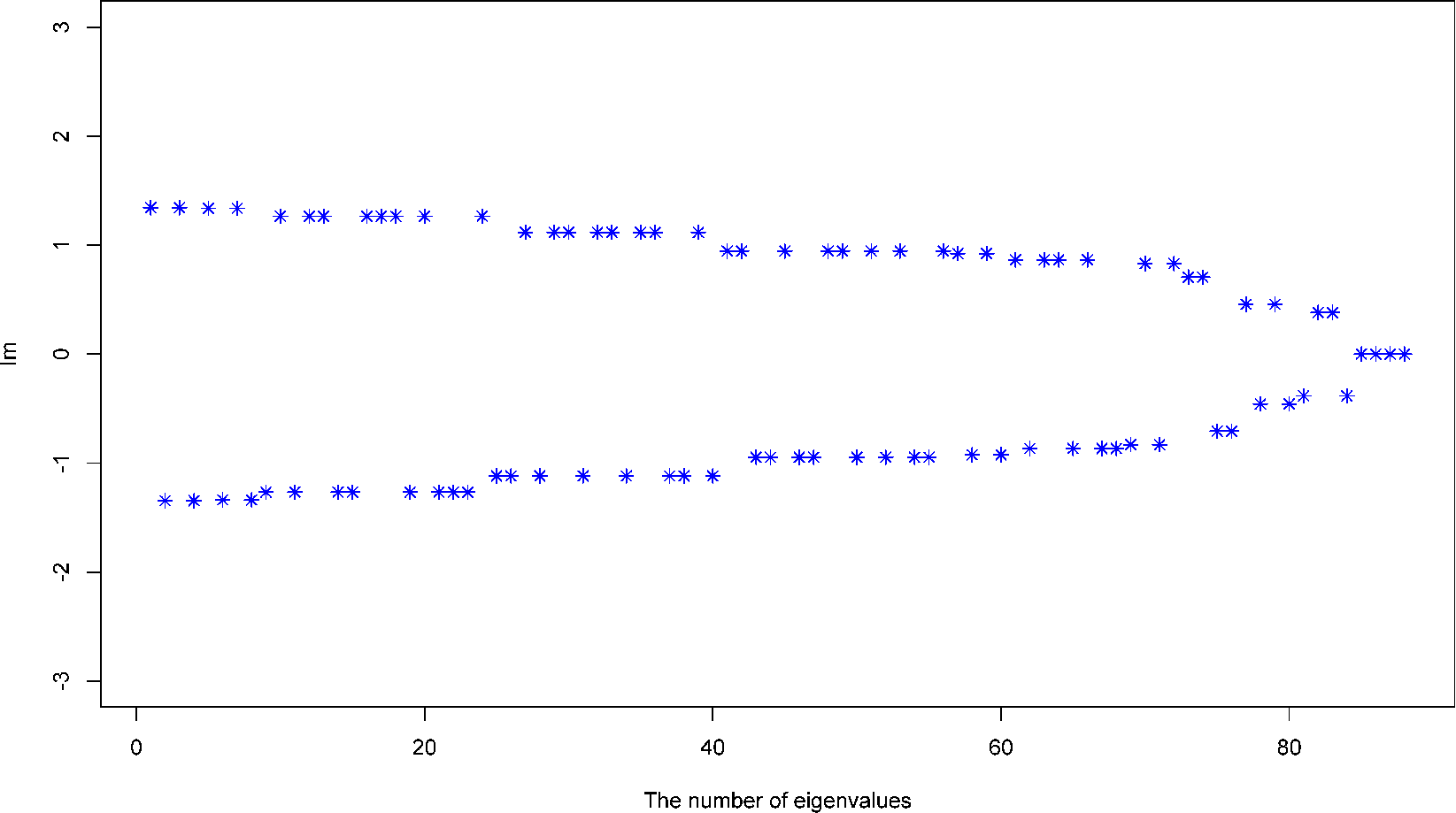}
\caption{Eigenvalue distribution is depicted, where the vertical axis represents the imaginary part of the Dirac matrix $\mathcal{D}^{(6,9)}$ and horizontal axis represents the serial number of eigenvalues.
There are four zero-eigenvalues corresponding to two Dirac zero-modes in two dimensions.
}
\label{plot2:2sphere69}
\end{figure}
\end{itemize}

We summarize our results in Table.~\ref{table:2sphere}.
\begin{table}[h]
 \caption{Maximal number of Dirac zero-modes on 2-sphere $(M,N)$}
 \label{table:2sphere}
 \centering
  \begin{tabular}{c|c|c}
   \hline
   \quad the number of $M$ \quad
   &\quad the number of $N$ \quad
   &\quad maximal $\#$ of Dirac zero-modes \quad \\
   \hline \hline
   even &even   &0\\
   \hline
   odd  &even   &0\\
   \hline
   odd  &odd    &1\\
   \hline
   even &odd    &2
  \end{tabular}
\end{table}


\subsection{Four-dimensional sphere}
\label{app:sphere2}

We discuss the discretized four-dimensional sphere.
We first consider the discretized four-dimensional spherical-coordinate system for $4$-sphere labeled by four integers $(N_{1},N_{2},N_{3},N_{4})$ as
\begin{gather}
    x_{5} = r\cos \theta_{4}, \qquad
	x_{4} = r\sin \theta_{4} \cos \theta_{3},\qquad
	x_{3} = r\sin \theta_{4} \sin \theta_{3} \cos \theta_{2},\\
	x_{2} = r\sin \theta_{4} \sin \theta_{3} \sin \theta_{2} \cos \theta_{1},\qquad
	x_{1} = r\sin \theta_{4} \sin \theta_{3} \sin \theta_{2} \sin \theta_{1}
\end{gather}
where $r$ is a radial distance and the four angles are discretized as $\theta_{1} \equiv \frac{2\pi}{N_{1}} \left( n_{1} - 1 \right),\ \theta_{i} \equiv \frac{\pi}{N_{i}-1} \left( n_{i} - 1 \right)$ for $i=2,3,4$. 
$n_{1},n_{i} \in \mathbb{N}$ run as $n_{1} \in [1,N_{1}],\ n_{i} \in [1,N_{i}]$.
For simplicity, we fix a radial distance as $r=1$.
In a parallel manner to the discussion for 2-sphere, we label the lattice sites as $(n_{1},n_{2},n_{3},n_{4})$.

\begin{figure}[htpb]
\centering
	\includegraphics[clip,
        height=4cm
		]{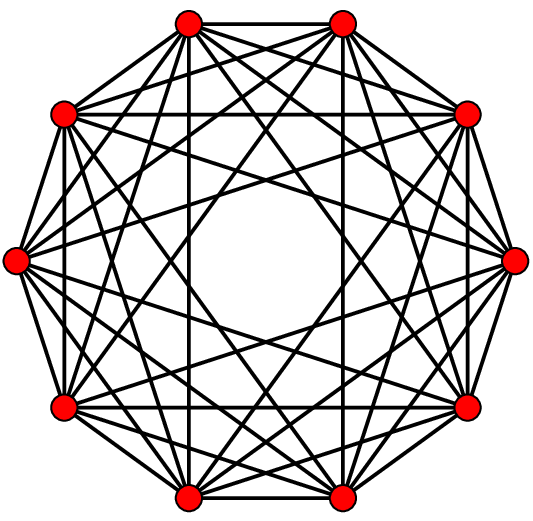}
\caption{
Graph corresponding to the $(4,3,3,3)$ 4-sphere is the 5-orthoplex inside Petrie polygon.
}
\label{graph:4sphere}
\end{figure}

For instance, we take $N_{1} = 4$ and $N_{i} = 3$ for $i=2,3,4$.
A graph corresponding to the $(4,3,3,3)$ 4-sphere is depicted in Fig.~\ref{graph:4sphere}.
The Dirac matrix is given as
\begin{equation}
\begin{split}
    \mathcal{D}^{(4,3,3,3)} &= 
    \begin{pmatrix}
        D_{4}\\
        &O_{6}
    \end{pmatrix}\otimes \gamma_{1}
    +
    \begin{pmatrix}
        O_{4}   &V_{4,2}\\
        -V_{4,2}^{\dagger}  &O_{2}\\
        &&O_{4}
    \end{pmatrix}\otimes \gamma_{2}\\
    &\quad
    +
    \begin{pmatrix}
        O_{6}   &V_{6,2}\\
        -V_{6,2}^{\dagger}  &O_{2}\\
        &&O_{2}
    \end{pmatrix}\otimes \gamma_{3}
    +
    \begin{pmatrix}
        O_{8}   &V_{8,2}\\
        -V_{8,2}^{\dagger}  &O_{2}
    \end{pmatrix}\otimes \gamma_{4}
\end{split}\,.
\end{equation}
This Dirac matrix $\mathcal{D}^{(4,3,3,3)}$ is $10 \times 10$ square matrix, apart from the $\gamma$ matrix structure.
\begin{figure}[htpb]
\centering
	\includegraphics[clip,
		width=0.8\linewidth
		]{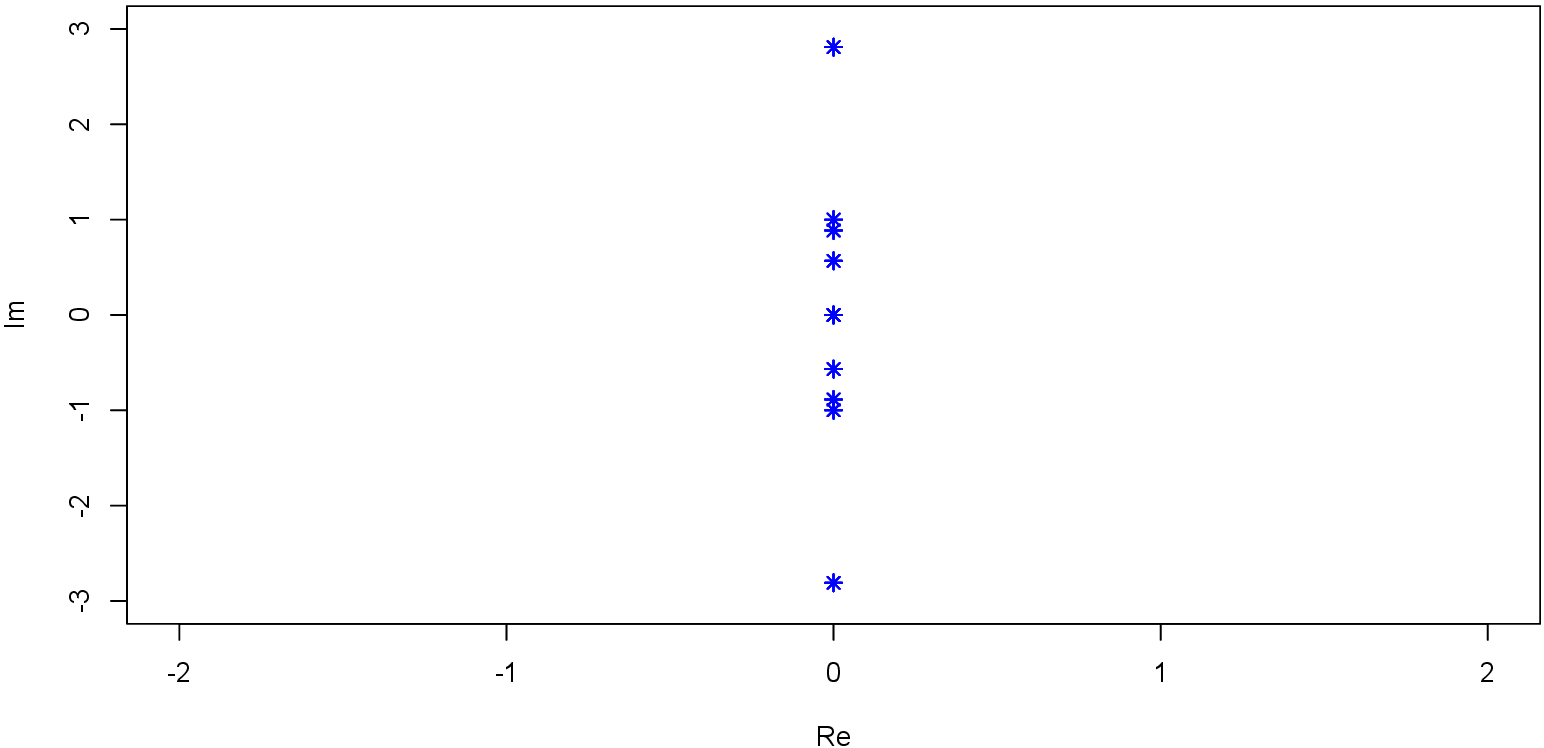}
\caption{Eigenvalue distribution of the Dirac matrix $\mathcal{D}^{(4,3,3,3)}$.
There are two Dirac zero-modes in four dimensions, which emerge as eight zero-eigenvalues in the figure.
}
\label{plot:4sphere}
\end{figure}
\begin{figure}[htpb]
\centering
	\includegraphics[clip,
		width=0.8\linewidth
		]{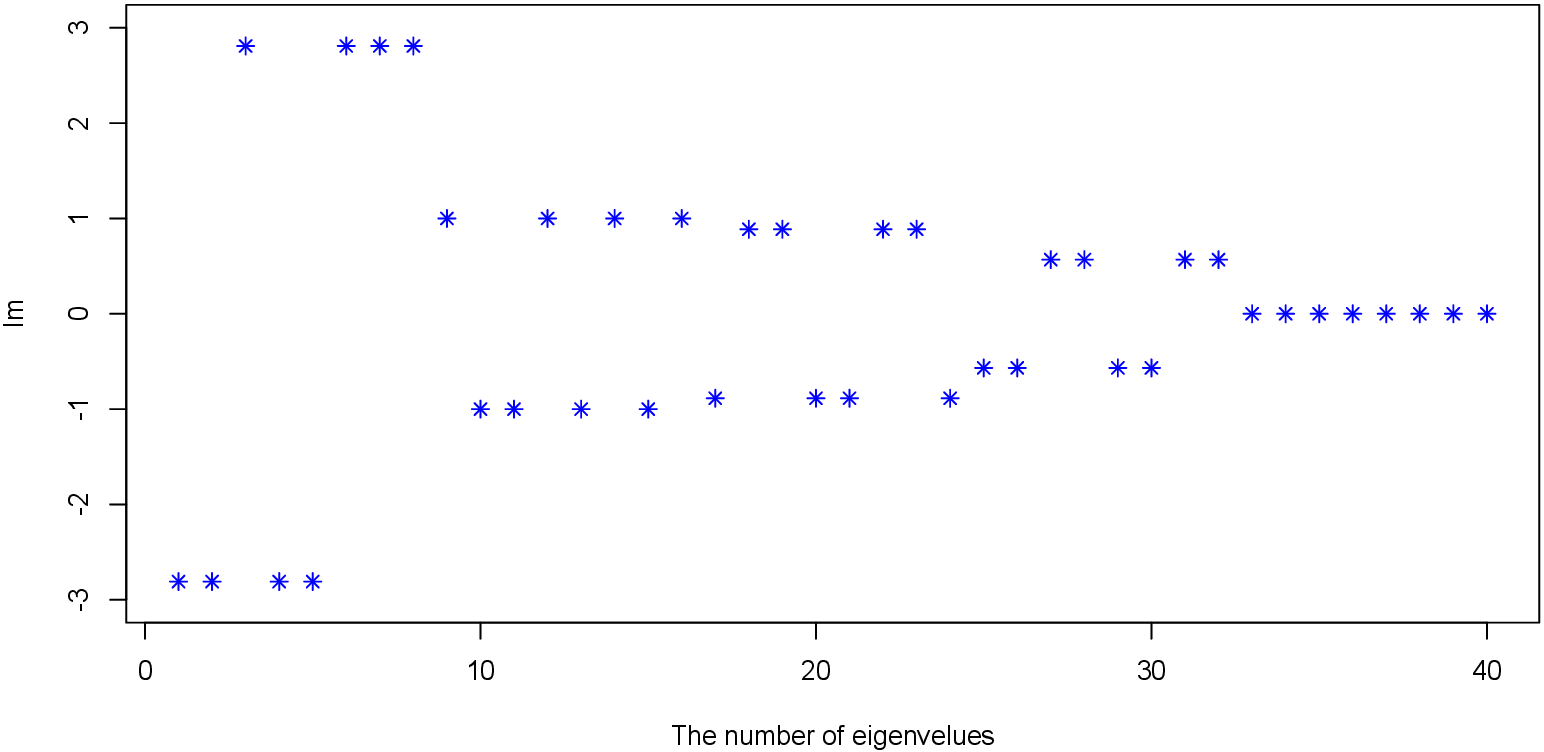}
\caption{Eigenvalues of the Dirac matrix $\mathcal{D}^{(4,3,3,3)}$.
There are eight zero-eigenvalues corresponding to two Dirac zero-modes in four dimensions.
}
\label{plot2:4sphere}
\end{figure}
Fig.~\ref{plot:4sphere} and Fig.~\ref{plot2:4sphere} shows the eigenvalue distributions of the Dirac matrix $\mathcal{D}^{(4,3,3,3)}$.
For the dicretized 4-sphere $(4,3,3,3)$, we find that there are two Dirac zero-modes as there are eight zero-eigenvalues of the Dirac matrix $\mathcal{D}^{(4,3,3,3)}$.

By studying other cases, we find that the number of Dirac zero-modes on the discretized 4-sphere $(N_{1},N_{2},N_{3},N_{4})$ is two when the number of sites $N_{1}$ is even and the number of sites $N_{i}$ for $i$ from $2$ to $4$ are odd, as with the case on the discretized 2-sphere $(M,N)$.
We could not find any example where the number of Dirac zero-modes goes beyond two in four dimensions too.
%


\bibliographystyle{utphys}
\bibliography{./QFT,./refs,./math}

\end{document}